\documentclass[format=acmlarge, screen=true, review=False, anonymous=False]{acmart}
\AtBeginDocument{%
  }

\setcopyright{acmlicensed}
\acmJournal{IMWUT}
\copyrightyear{2025}
\acmYear{2025}
\acmDOI{XXXXXXX.XXXXXXX}
\acmISBN{978-1-4503-XXXX-X/2018/06}

\usepackage{xcolor}
\usepackage{mdframed}
\newcommand{\highlight}[1]{\vspace{0.0cm}\begin{mdframed}[backgroundcolor=gray!15]#1\end{mdframed}}
\mdfsetup{skipabove=1em,skipbelow=1em}

\usepackage{rotating}
\usepackage{adjustbox}

\usepackage{arydshln}
\let\savedhline\hline
\let\savedcline\cline

\usepackage{longtable}
\let\hline\savedhline
\let\cline\savedcline

\usepackage{booktabs}
\usepackage{caption}
\usepackage{graphicx}
\usepackage{subcaption}

\usepackage{hyperref}
\usepackage{multirow}
\usepackage{multicol}
\usepackage{color}
\usepackage{colortbl}
\definecolor{Gray}{gray}{0.9}
\newcolumntype{g}{>{\columncolor{Gray}}c}
\usepackage{graphicx}
\usepackage{makecell}

\newcommand{\add}[1]{\textcolor{black}{#1}}
\newcommand{\delete}[1]{}
\newcommand{\modif}[1]{\textcolor{black}{#1}}

\begin{document}


\title{\modif{A Review of Behavioral Closed-Loop Paradigm from Sensing to Intervention for Ingestion Health}}

\author{Jun Fang}
\authornote{Both authors contributed equally to this research.}
\email{fangy23@mails.tsinghua.edu.cn}
\orcid{0009-0001-2614-8674}
\author{Yanuo Zhou}
\authornotemark[1]
\email{zhou-yn24@mails.tsinghua.edu.cn}
\orcid{0009-0009-0652-2196}
\affiliation{%
  \institution{Department of Computer Science and Technology, Tsinghua University}
  \country{China}
}

\author{Ka I Chan}
\affiliation{
  \institution{School of Information, University of Michigan}
  \city{Ann Arbor}
  \country{United States}}
\email{chankai@umich.edu}
\orcid{0009-0001-4560-1702}

\author{Jiajin Li}
\authornote{Visiting student at the time of this work.}
\email{lijiajin0516@gmail.com}
\affiliation{
  \institution{Tsinghua University}
  \country{China}}
\orcid{0009-0001-5723-5383}
\email{}

\author{Zeyi Sun}
\orcid{0009-0001-8358-2450}
\affiliation{
   \institution{Tsinghua University}
  \country{China}}
\email{lavenderine123@gmail.com}

\author{Zhengnan Li}
\orcid{0009-0008-0592-214X}
\affiliation{
  \institution{Communication University of China}
  \country{China}}
\email{lzhengnan389@gmail.com}

\author{Zicong Fu}
\orcid{0009-0003-4566-6009}
\email{fuzc23@mails.tsinghua.edu.cn}
\author{Hongjing Piao}
\email{phj23@mails.tsinghua.edu.cn}
\orcid{0009-0008-5690-7353}
\author{Haodong Xu}
\orcid{0009-0003-4283-648X}
\email{xhd23@mails.tsinghua.edu.cn}
\affiliation{%
  \institution{Tsinghua University}
  \country{China}
}

\author{Yuanchun Shi}
\orcid{0000-0003-2273-6927}
\affiliation{%
  \institution{Key Laboratory of Pervasive Computing, Ministry of Education, Department of Computer Science and Technology, Tsinghua University}
  \country{China}
}
\affiliation{%
  \institution{Intelligent Computing and Application Laboratory of Qinghai Province, Qinghai University}
  \country{China}
}
\email{shiyc@tsinghua.edu.cn}

\author{Yuntao Wang}
\authornote{Corresponding authors.}
\email{yuntaowang@tsinghua.edu.cn}
\orcid{0000-0002-4249-8893}
\affiliation{%
  \institution{Key Laboratory of Pervasive Computing, Ministry of Education, Department of Computer Science and Technology, Tsinghua University}
  \country{China}
}

\renewcommand{\shortauthors}{Fang, Zhou, et al.}

\begin{abstract}

\begin{figure}[h]
    \centering
    \includegraphics[width=\linewidth]{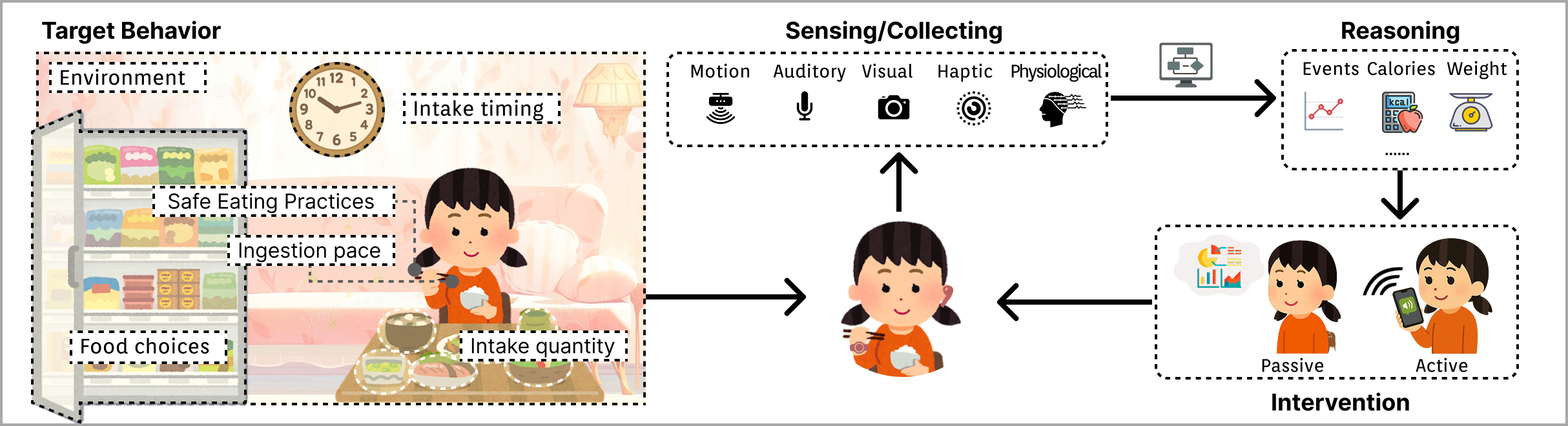}
    \caption{\modif{Paradigm of ingestion health from sensing to intervention}}
    \label{fig:framework_fig}
\end{figure}

Ingestive behavior plays a critical role in health, yet many existing interventions remain limited to static guidance or manual self-tracking. With the increasing integration of sensors\modif{, context-aware computing,} and perceptual computing, recent systems have begun to support closed-loop interventions that dynamically sense user behavior and provide feedback during or around ingestion episodes. In this survey, we review 136 studies that leverage sensor-enabled or interaction-mediated approaches to influence \modif{ingestive} behavior. We propose a behavioral closed-loop paradigm \add{rooted in context-aware computing and inspired by HCI behavior change frameworks,} comprising \modif{four} components: target behaviors, sensing modalities, \modif{reasoning} and \modif{intervention} strategies. A taxonomy of sensing and intervention modalities is presented, organized along human- and environment-based dimensions. Our analysis also examines evaluation methods and design trends across different modality-behavior pairings. This review reveals prevailing patterns and critical gaps, offering design insights for future adaptive and context-aware ingestion health interventions.

\end{abstract}




\maketitle

\section{Introduction}
\label{intro}

Ingestion health \modif{plays a critical role in shaping health outcomes, influencing} quality of life, longevity, and overall well-being \cite{jovanov2014sensors, mesas2012selected}. \modif{Despite its importance, current ingestion health interventions often rely on static tools} such as food diaries \cite{thompson2017dietary}, self-report questionnaires \cite{shim2014dietary}, and periodic counseling sessions \cite{greaves2011systematic}, \modif{which place a heavy burden on users and provide limited behavioral feedback \cite{raber2021systematic, takemoto2018diet}.} As a result, they often fail to dynamically support individuals in developing sustainable and healthy ingestive behaviors over time.

\modif{The emergence of ubiquitous computing has long envisioned systems that seamlessly integrate into everyday life, sensing and responding to context to enhance human activity \cite{weiser1991computer}. Advances in sensing technologies have further enabled low-burden, continuous monitoring of ingestive behaviors. In particular, multimodal sensor fusion has made unobtrusive ingestion monitoring feasible \cite{bell2020automatic}, allowing for the collection of rich behavioral data with minimal user effort.}


\modif{Context-aware computing is a core capability of ubiquitous computing \cite{abowd2000charting}, and it advocates modular architectures in which continuous sensing informs higher-level reasoning to enable adaptive interventions. \citet{dey2001understanding} defined context-aware systems as those that \textit{"use context to provide relevant services to the user, where relevancy depends on the user’s task"}. Accordingly, sensing is not merely about acquiring raw data, but about capturing context to enable situation inference. Intervention, then, corresponds to the application component that, as \citet{abowd2000charting} emphasized, makes decisions and takes actions based on inferred context. These principles support personalized, adaptive interventions grounded in real-time behavioral context \cite{bell2020automatic}, marking a shift from static, user-driven methods to context-aware, behavior-informed systems.} 

\modif{Studies of ingestion health related to the aforementioned context have been systematically reviewed, with prior reviews primarily focusing on three key aspects}: dietary assessment, wearable ingestion detection, and behavior change intervention.

\add{\textbf{i.}} \textbf{Dietary Assessment:}
\modif{\citet{skinner2020future} and \citet{wang2022enhancing} reviewed sensor-based frameworks for improving dietary assessment accuracy. However, their scope remains limited to intake monitoring, without addressing how such data can inform adaptive interventions.}

\add{\textbf{ii.}} \textbf{Wearable Ingestion Detection:}
\modif{Reviews by \citet{bell2020automatic} and \citet{he2020comprehensive} systematically analyzed sensor types and detection pipelines, while others focused on chewing-specific sensing \cite{wei2022review, mortazavi2023review}. These works emphasize detection performance but rarely consider downstream behavior change frameworks.}

\add{\textbf{iii.}} \textbf{Behavior Change Intervention:}
\modif{\citet{zhang2020design} and \citet{pan2021digital} reviewed digital tools for eating-related interventions but primarily centered on static strategies such as education or gamification, with limited integration of sensing or context-aware intervention.}

\modif{In summary, prior reviews tend to isolate sensing, reasoning, or intervention, without linking them into an integrated behavior change pipeline. This highlights the need for a systematic review that explicitly integrates ingestion health interventions and perceptual computing into a cohesive system. To address this, we propose a behavioral closed-loop paradigm for ingestion health that encompasses sensing, reasoning, and intervention. This paradigm builds on Salber’s Context Toolkit \cite{salber1999context}, inspired by Personal Informatics Framework \cite{li2010stage} and Digital Behavior Change Intervention \cite{hekler2016advancing}, aiming to bridge the scattered components identified in previous reviews. Moreover, it advances from passive context display toward more active and adaptive forms of behavioral scaffolding.}

\modif{Building on this conceptual foundation,} we conduct a structured review guided by the following research questions (RQs), aimed at evaluating how ingestion health interventions are conceptualized and implemented within behavioral closed-loop systems:
\begin{itemize}
    \item RQ1: What behaviors are targeted in the existing literature on ingestion health intervention?
    \item RQ2: What are the structural characteristics and design space of perceptual computing-based paradigms for sensing and intervening in ingestive behavior?
    \item RQ3: How effective are current intervention paradigms, and what outcomes have they achieved?
    \item RQ4: What are the gaps in the implementation of behavioral closed-loop systems in current ingestion health interventions, and what are the recommendations for them?
\end{itemize}

\textit{Contribution Statement}: 

\begin{itemize}
\item We conduct a systematic review of ingestion health interventions that employ behavioral sensing and interactive systems, and develop a taxonomy of key components across different stages of the behavioral closed-loop paradigm.
\item We propose a behavioral closed-loop \modif{paradigm, grounded in context-aware computing and informed by HCI behavior change frameworks, to guide} ingestive behavior interventions, and review the evaluation of effectiveness and user acceptance in current literature.
\item We identify critical gaps between current ingestion health interventions and real-world behavioral complexities, and provide design insights for better bridging behavioral targets, sensing, reasoning, and intervention stages toward effective closed-loop management.
\end{itemize}

\section{Method}
\label{method}

Drawing inspiration from prior review methodologies \cite{zhang2020design, jansen2022design, motahar2022review}, we conducted a \textit{Systematic Literature Review (SLR)} to address our research questions. Specifically, our review process followed the guidelines of the \textit{Preferred Reporting Items for Systematic Reviews and Meta-Analyses (PRISMA)} proposed by \citet{page2021prisma}.

\subsection{Paper Retrieval}

Given that most relevant publications place greater emphasis on the\delete{ feedback or} intervention aspect, we also include studies that do not cover the full pipeline from sensing to intervention. After careful discussion, we decided to include all publications in which sensor-based interactive devices or perceptual computing methods are involved at any stage within the behavioral closed-loop process, ranging from sensing to\delete{ feedback or} intervention.

To ensure consistency in inclusion and classification throughout in this paper, we first examined three core terms and finalized their definitions as follows:

 \begin{itemize}
     \item Ingestive Behavior: The process by which an \textbf{individual} ingests \textbf{nutrition} from food or liquid intake, excluding the intake of medications or medical nutrition supplements, and comprising appetitive behaviors, consummatory behaviors, and post-ingestive physiological responses;
     \item Health Intervention: \modif{A system-generated mechanism aimed at \textbf{altering} user behavior for health improvement, categorized as \textbf{passive interventions} --- providing \textbf{feedback} to foster \textbf{self-awareness}, and \textbf{active interventions} --- delivering \textbf{structured guidance} or \textbf{direct mechanisms} to proactively guide behavior.}
     \item Perceptual Computing: Technologies that enable machines to \textbf{perceive}, \textbf{compute}, and \textbf{reason} with data from \textbf{real-world sensory inputs}.
 \end{itemize}

\add{While the paradigm is conceptually grounded in context-aware computing, we adopt "perceptual computing" as a core operational term. This term provides a more implementation-oriented perspective on behavioral data processing in closed-loop systems building on context-aware principles. It focuses on gaps in prior reviews where perception and intervention were often treated as disconnected components.}

Specifically, \citet{benoit2008behavioral} defined ingestive behavior as "\textit{a complex process controlled by multiple physiological and psychological mechanisms}", and divided it \modif{into} two general categories, including "\textit{behaviors involved in seeking, locating, and acquiring foods}" and "\textit{behaviors associated with the act of ingestion itself, including chewing and swallowing}". We refer to these two categories as "appetitive behaviors" and "consummatory behaviors", and additionally added "post-ingestive behaviors" to cover physiological feedback mechanisms and related behaviors after ingestive behavior. Moreover, \citet{mendel2010perceptual} described perceptual computing as a sensor-based method "\textit{associated with machinery for computing and reasoning with perceptions and data}", while \citet{poole2013hci} noted that "\textit{health interventions leverage technology to deliver timely support, monitoring, and guidance to influence health behaviors and outcomes}". These definitions align closely with the conceptual scope of our study. Accordingly, we adopted and refined the above terms to form the foundation of our inclusion criteria and analytical framework.

Following the PRISMA principle and the search strategy adopted in related work \cite{hiraguchi2023technology}, we began our process by identifying keywords terms closely related to core definitions. Moreover, to refine the keyword strategy for literature retrieval, we compiled a seed set of approximately 50 papers, including some reviewed studies that clearly met our inclusion criteria, and top-ranked search results from digital libraries \cite{roddiger2022sensing}. Subsequently, we applied a TF-IDF \cite{salton1975vector} and N-gram \cite{cavnar1994n} analysis on the titles, abstracts, and keywords of this seed set to identify additional high-frequency or semantically relevant terms. These were used to augment our original keyword list, ensuring broader coverage of technically and behaviorally relevant studies. To illustrate the outcome of this refinement process, we present the top five high-frequency terms extracted via TF-IDF and Bi-gram analysis (excluding semantically uninformative combinations such as function words or prepositional phrases) as follows:

\highlight{
\textbf{TF-IDF}: eating, behavior, feedback, food, intake

\noindent{
\textbf{Bi-gram}: behavior(al) change, eating rate, eating habits, eating behavior, to improve
}
}

We selected three primary databases for our literature retrieval: the ACM Digital Library, IEEE Xplore, and Google Scholar. These sources are considered to comprehensively cover publications closely related to sensing technologies, interactive devices, and the Human–Computer Interaction (HCI) domain.
Given the variation in keyword syntax and thesaurus constraints across platforms, we tailored the search strings and scoping targets for each database accordingly. The final keyword formulations \modif{are summarized below, with corresponding search details for each source shown in Appendix \ref{search_results}}. All searches were conducted on April 4, 2025, without restricting the publication time interval.

\highlight{
\textbf{Keywords (combined with AND)}: 
\begin{itemize}
    \item eating-related: eat*, "food", diet*, "intake"
    \item intervention-related: interven*, "feedback", "behavior(al) change", promot*, adjust*, design*, improv*, alter*
    \item technology-related: "wearable", "sensor(s)", "device(s)", "interactive"
\end{itemize}
}

\subsection{Screening}
\label{chapter:selected_papers}

To further narrow down the set of retrieved articles to those aligned with the defined scope of our review, two authors conducted a structured two-stage screening process \delete{in accordance }with PRISMA guidelines. During the initial screening, which involved title and abstract review followed by full-text assessment, each article was evaluated against a predefined set of inclusion and exclusion criteria. Articles were labeled as "include", "exclude", or "uncertain" and any disagreements were resolved through discussion until consensus was reached. The selection criteria used for this process are outlined below and have been adapted to follow PRISMA recommendations.

(1) The paper was not accessible in full text or was not written in English;  

(2) The paper was a duplicate or an extended version of a previously included record;  

(3) The study focused on behaviors outside our scope or not pertaining to ingestive behavior;

(4) The study did not include any behavioral intervention in the process (e.g., detection-only systems);  

(5) The system did not involve any sensor, interactive device, or perceptual computing capability;  

(6) The study targeted clinical or therapeutic contexts (e.g., medication ingestion, artificial feeding);  

(7) The intervention goal was not health-related (e.g., flavor or entertainment enhancement);  

(8) The paper was not a primary research article (e.g., reviews, position papers, dataset descriptions, fictional designs).

Two authors reviewed 1690 relevant publications (679 ACM-DL, 911 IEEE-X, 100 Google Scholar) based on the selecting criteria and confirmed 98 papers (50 ACM-DL, 40 IEEE-X, 8 Google Scholar) for our downstream analysis. Then, we added 9 papers from our previous bibliography and conducted backward-chaining to fill in some papers published in other databases or neglected from keywords. An initial set of papers from ACM and IEEE collections was screened by examining related work sections, and the selections were independently reviewed by another author. 29 additional papers were selected via backward-chain, and among this 8 papers are from IEEE and 7 papers are from ACM, whose missions are due to keywords. Moreover, 8 papers are from Springer, 3 papers are from NIH and other 3 papers are from ResearchGate.

\begin{figure}[htbp]
    \centering
    \includegraphics[width=\linewidth]{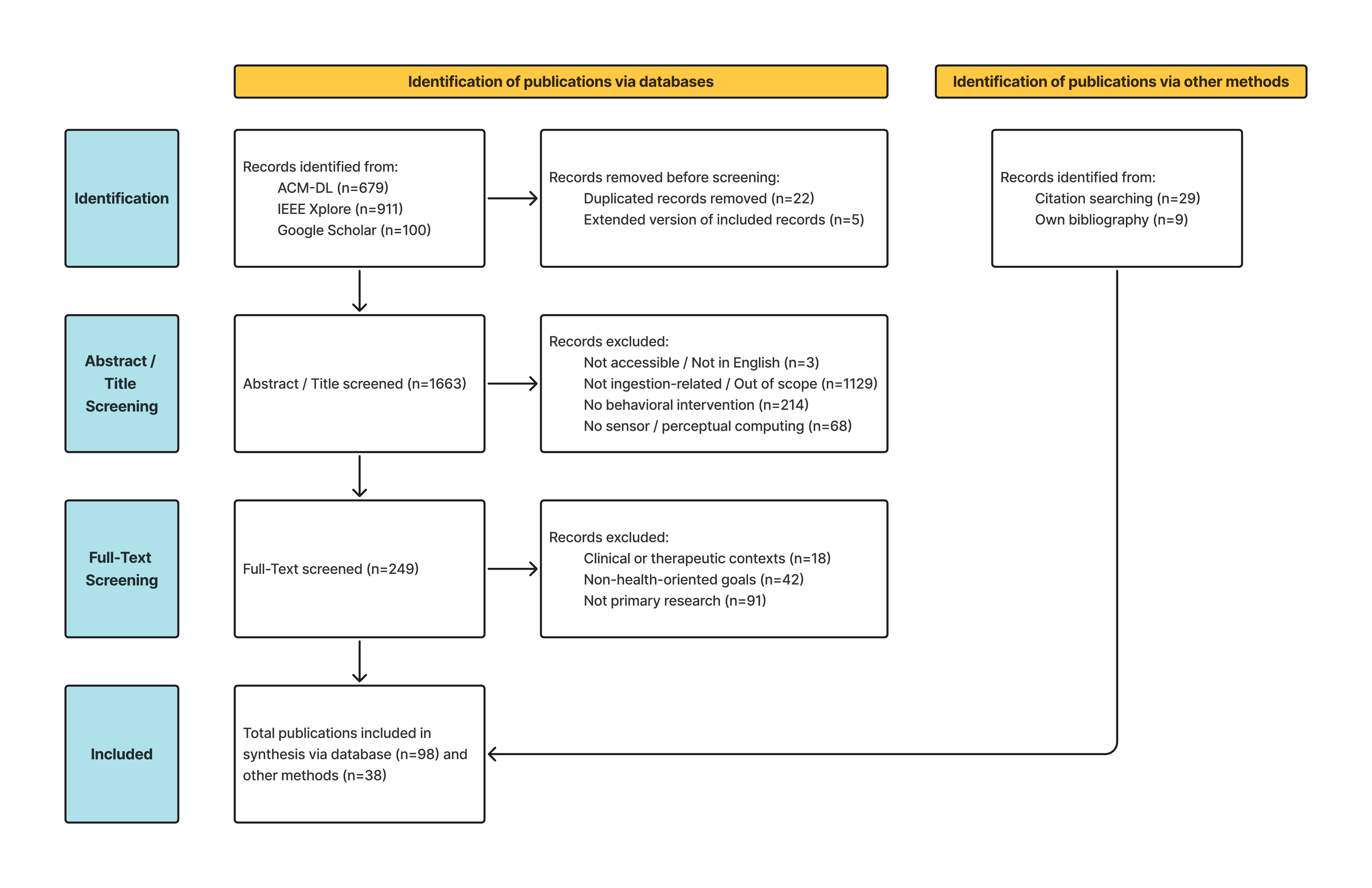}
    \caption{Flow chart of SLR process based on the PRISMA guidelines}
    \label{fig:flowchart}
\end{figure}

\begin{figure}[htbp]
    \centering
    \includegraphics[width=\linewidth]{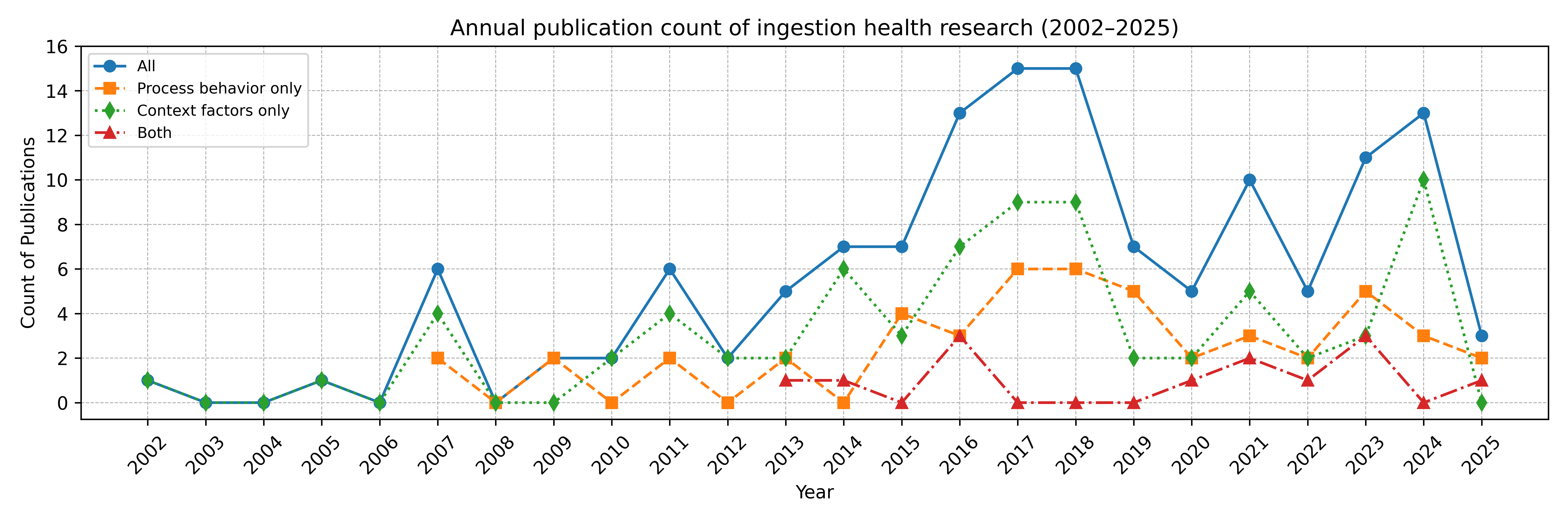}
    \caption{Total number of papers published overall (split to process behavior only, context factor only, and both included)}
    \label{fig:annual_count}
\end{figure}

Our overall multi-staged screening and selection process is depicted in Figure \ref{fig:flowchart}. The final set of review paper set includes 136 papers overall. Annual publication count is shown in Figure \ref{fig:annual_count}, illustrating a growing research interest in ingestion health in recent years. We identified 93 distinct publication venues across all included studies, with full details provided in \add{Table} \ref{tab:venues list}. Among them, the following venues featured two or more included papers: ACM Conference on Human Factors in Computing Systems (CHI, n=16), ACM on Interactive, Mobile, Wearable and Ubiquitous Technologies (IMWUT)/UbiComp (N=16), ACM Interaction Design and Children Conference (IDC, n=2), ACM Designing Interactive Systems Conference (DIS, n=4), IEEE Sensors (n=2), IEEE International Conference on Robot and Human Interactive Communication (RO-MAN, n=2), ACM/IEEE International Conference on Human-Robot Interaction (HRI, n=2), PervasiveHealth (n=2), SIGGRAPH Asia (n=2), International Conference on Tangible, Embedded, and Embodied Interaction (TEI, n=2), Innovations in Power and Advanced Computing Technologies (i-PACT, n=2), International Conference on Computing Communication and Networking Technologies (ICCCNT, n=2), World Congress on Medical Physics and Biomedical Engineering (n=2).

To facilitate systematic analysis, we created a Google Sheets table with 57 columns. \modif{The column structure was iteratively developed to support the discovery and refinement of the taxonomy of core features (corresponding to Chapter~\ref{chapter:rq1} and Chapter~\ref{chapter:rq2}).} Two authors first defined preliminary categories during the filtering stage, followed by collaborative refinement of the taxonomy through a full review of the papers. Afterward, papers were split among authors for detailed annotation, with further refinements made throughout the reading process.

\section{Result Overview}

Following the review methodology and \delete{the }research questions outlined above, this section presents a two-part overview of \modif{papers selected in Section \ref{chapter:selected_papers}}. \modif{We first classify the publications based on Just-in-Time (JIT) characteristics, highlighting their temporal and contextual responsiveness. Then, we introduce our behavioral closed-loop paradigm grounded in context-aware computing and inspired by HCI behavior change frameworks.} This paradigm outlines key stages from sensing to intervention and serves as the analytical foundation for the subsequent taxonomy and discussion.

\subsection{Primary Taxonomy based on JIT}

\modif{Building on prior work emphasizing the importance of timing and context-awareness in behavioral interventions \cite{hsu2025personalized},} we classified the included studies based on whether their intervention strategies exhibit JIT characteristics. \delete{With a primary overview based on JIT, }\modif{This} classification offers a high-level understanding of the integration mechanisms across existing \modif{ingestion health systems}, encompassing both sensing and \delete{feedback/}intervention components. 

\begin{figure}[htbp]
  \centering
  \subcaptionbox{Cumulative publication count\label{fig:annual_increasing_jit}}
    {\includegraphics[width=0.65\linewidth]{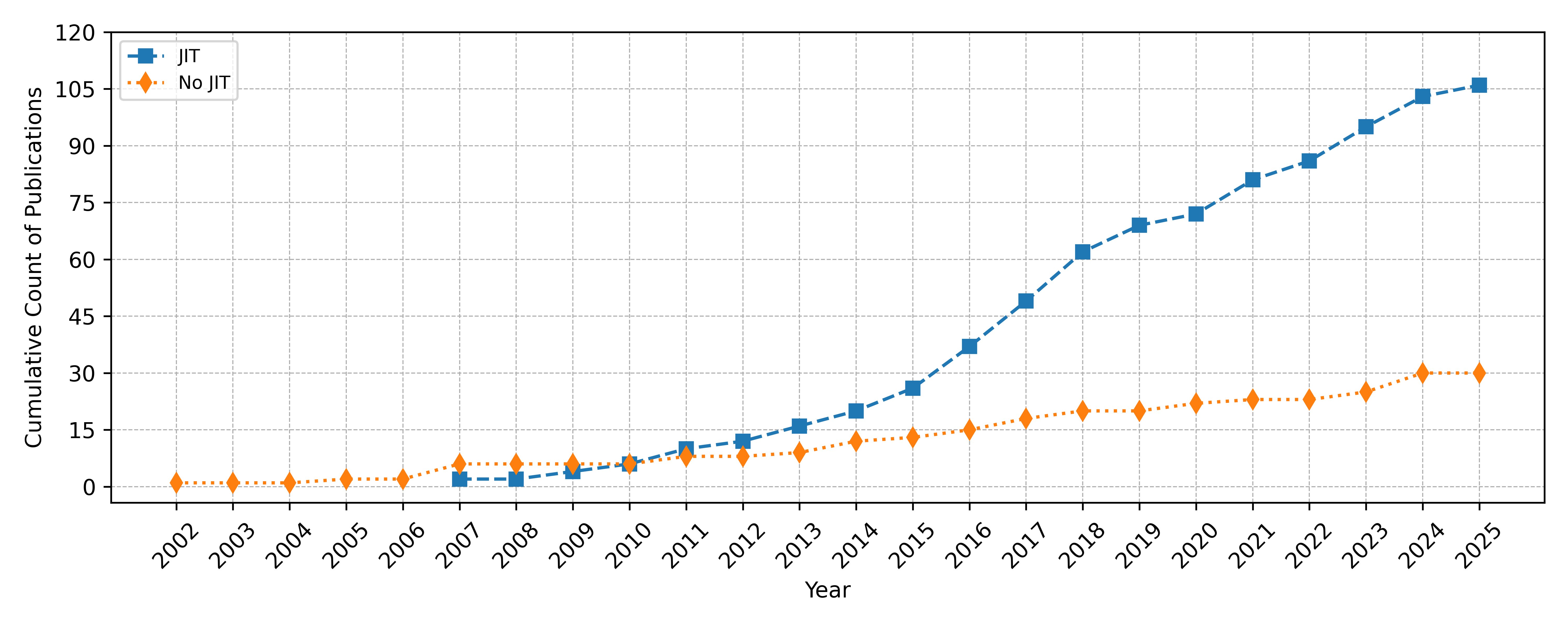}}
  \subcaptionbox{JIT - No JIT Pie\label{fig:jit_pit}}
    {\includegraphics[width=0.3\linewidth]{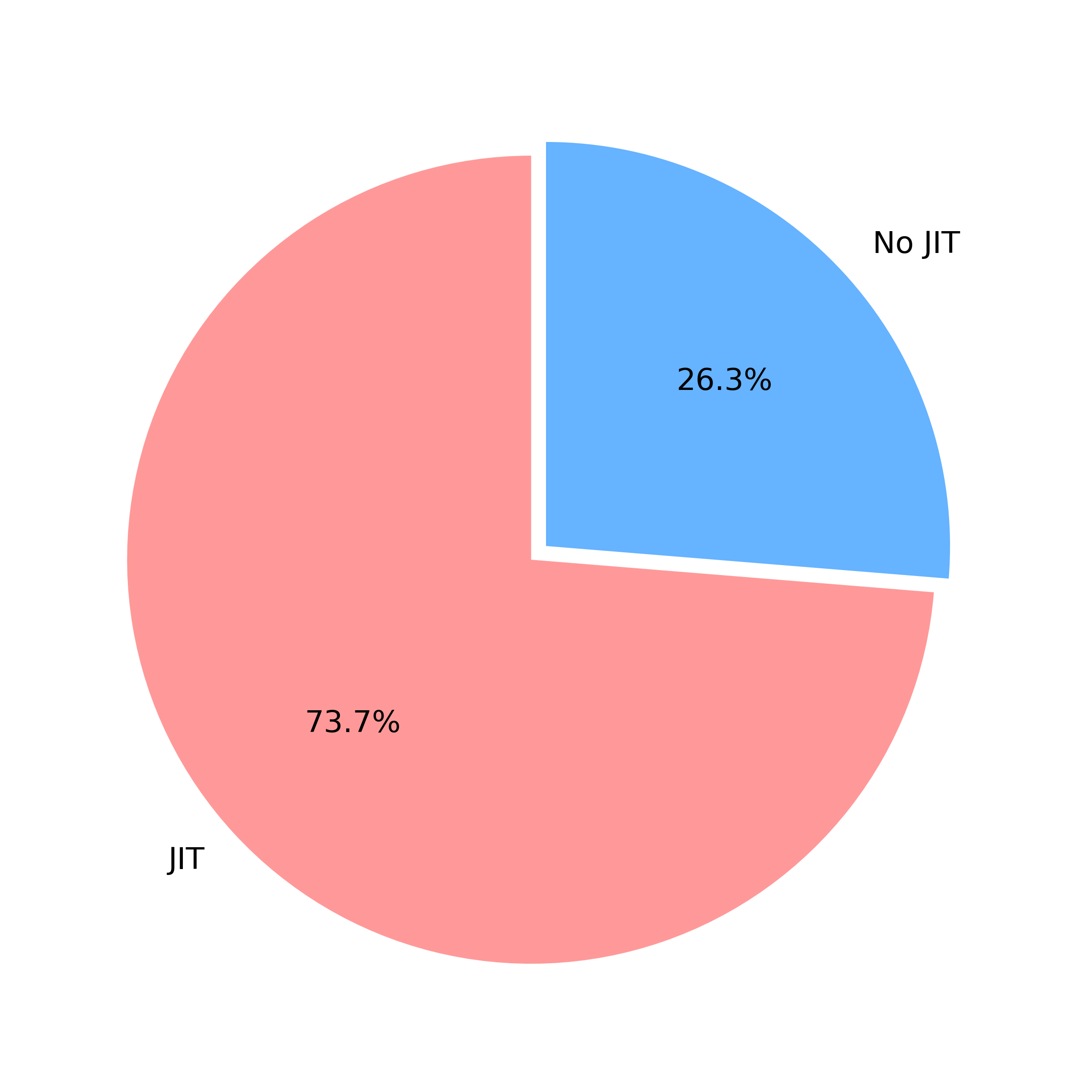}}
  \caption{Cumulative publication count of ingestion health research  (a) and publications pie (b).}
  \label{fig:JIT}
\end{figure}

JIT interventions are defined as approaches that deliver support dynamically at behaviorally or contextually opportune moments, \modif{ determined by} real-time assessments of users’ behavioral, physiological, or environmental states\modif{—particularly when individuals are vulnerable and receptive to intervention \cite{nahum2018just, sarker2014assessing, choi2019multi}}. \modif{As illustrated in Figure \ref{fig:annual_increasing_jit}, JIT interventions have gained increasing attention since their emergence around 2007, surpassing non-JIT studies by 2011. A notable surge occurred after 2015, driven by advances in wearable sensing and the availability of real-time contextual data. Overall, JIT interventions account for 73.7\% of the reviewed studies, as shown in Figure \ref{fig:jit_pit}, highlighting their central role in contemporary ingestion health intervention design.}

\subsection{Behavioral Closed-loop \modif{Paradigm}}
\label{overview_paradigm}

\modif{We propose a behavioral closed-loop paradigm that integrates various components --- sensing, reasoning, and intervention --- into a cohesive system. This paradigm is conceptually grounded in Human-Computer Interaction (HCI), drawing inspiration from context-aware computing and behavior change fields. Specifically, context-aware theories provide the theoretical foundation for system components, while HCI behavior change frameworks inform the behavioral logic of the closed-loop paradigm, resulting in an integrated closed-loop system.}

\subsubsection{\add{Grounded in Context-Aware Computing}}
\add{Context-aware computing emphasizes how interactive systems continuously sense, interpret, and respond to contextual information to enable adaptive, task-relevant interventions. Building on Dey’s foundational definition, context is defined as "\textit{any information that can be used to characterize the situation of an entity}",  and a context-aware system "\textit{uses context to provide relevant information and/or services to the user, where relevancy depends on the user’s task}" \cite{dey2001understanding}. These two definitions align with our paradigm's components in ingestion health systems: \textbf{Sensing} and \textbf{Reasoning}. Additionally, the Context Toolkit proposed by \citet{salber1999context} highlights a key principle: the separation of sensing, context interpretation, and application logic into modular components. Consequently, we have derived three main components in our paradigm:}
\add{\begin{itemize}
    \item \textbf{Sensing}: Raw sensor inputs are first collected as behavioral signals.
    \item \textbf{Reasoning}: Contextual signals are combined and interpreted to infer the user’s situation.
    \item \textbf{Intervention}: Application components make task-relevant decisions and deliver adaptive actions based on interpreted context.
\end{itemize}}

\subsubsection{\add{Inspired from Behavior Change Frameworks}}
\add{We then turn to behavioral change frameworks for a process-oriented perspective that clarifies how components can be organized temporally and functionally. The Personal Informatics Framework proposed by \citet{li2010stage} describes a five-stage cyclical process: preparation, collection, integration, reflection, and action. This framework inspired us to conceptualize our paradigm as a \textbf{behavioral loop}, mapping collection to \textbf{Sensing}, integration to \textbf{Reasoning}, and the reflection–action sequence to \textbf{Intervention}, framing the paradigm as a loop that enables continuous adjustment. Building on this, the Digital Behavior Change Interventions (DBCI) framework \cite{hekler2016advancing} provides system-level insights by integrating sensing, tailoring, action, and interaction into a unified architecture for adaptive interventions \cite{yardley2016understanding}. This approach led us to design the components to function not as isolated modules but as an integrated system.}

\add{While our paradigm focuses on three system-level components, it draws inspiration from the five-stage PIF process. Some stages, such as preparation, are embedded implicitly within broader sensing or reasoning phases.}

\subsubsection{\add{Strengthened with Closed-Loop Concept}}
\add{To reinforce the "loop" structure, we draw from broader "closed-loop" architectures in HCI. According to \citet{fairclough2015closed}, closed-loop systems continuously monitor user states and feed those signals into adaptive controllers to trigger context-appropriate responses. \citet{murray2024active} further conceptualize such systems as mutually interactive agents, where both the user and the system perceive and act on each other’s changing states over time.}

\begin{figure}[htbp]
    \centering
    \includegraphics[width=\linewidth]{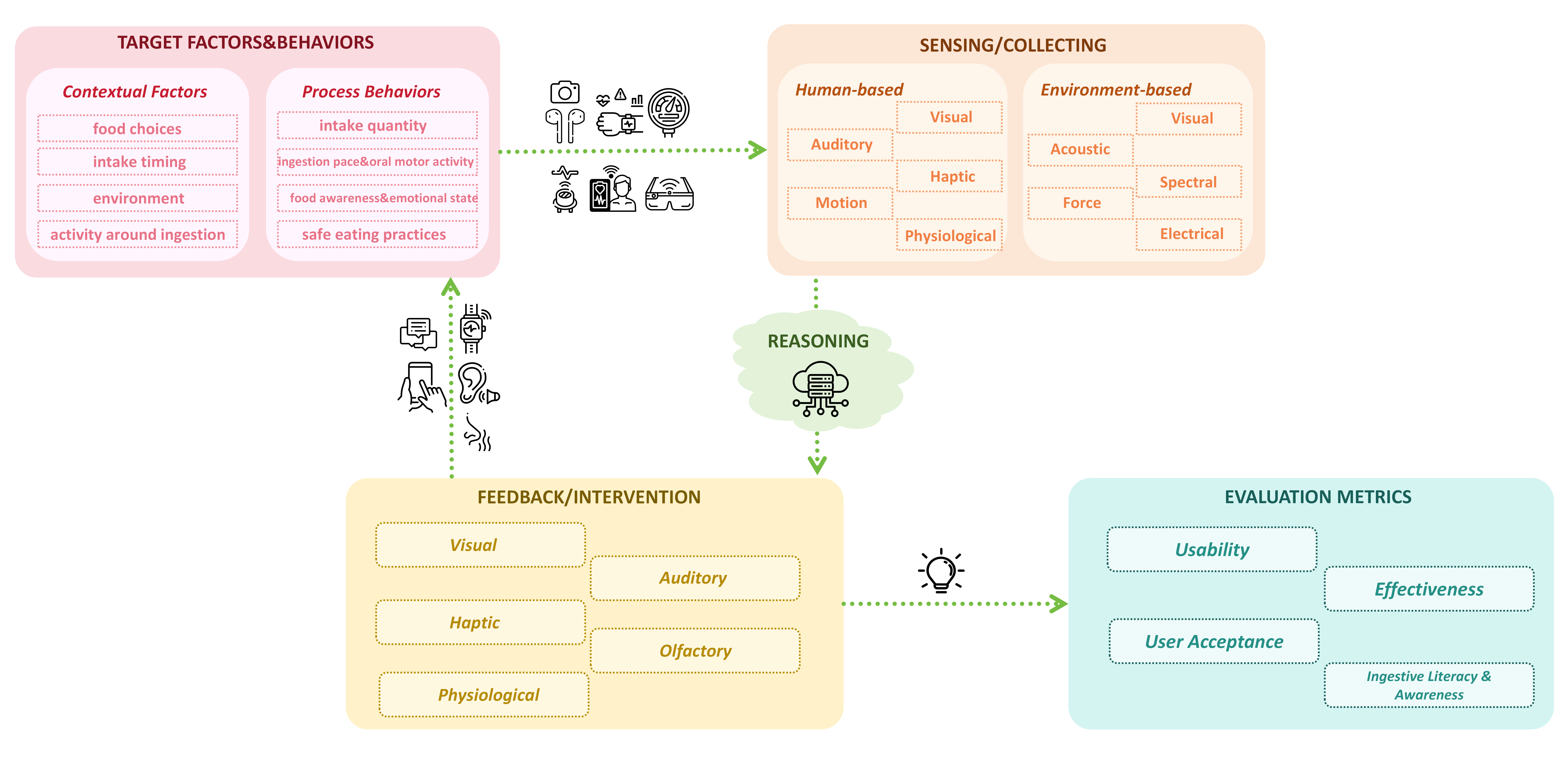}
    \caption{Behavior closed-loop overview}
    \label{fig:close-loop overview}
\end{figure}

\add{Drawing from these perspectives, we develop our behavioral closed-loop paradigm as a dynamic, sensing-driven system that continuously interprets user behavior and delivers adaptive interventions, enabling iterative feedback over time and context. This feedback can arise from real-time behavioral data, user-state inference, or performance evaluation, enabling adaptive recalibration across the loop. Instead of a monolithic process, this paradigm comprises recurring stages: identifying target behaviors, sensing relevant data, delivering interventions, and evaluating outcomes to guide future actions. When applied across everyday environments, these loops form dense, time- and context-sensitive patterns, realizing the vision of ubiquitous computing through distributed, context-aware interventions. Figure~\ref{fig:close-loop overview} illustrates the core structure and evolution of the behavioral closed-loop paradigm, informing our subsequent taxonomy and system-level analysis.}

\section{Taxonomy of Target Factors and Behaviors}
\label{chapter:rq1}

Daily ingestive behavior interventions can play an important role in promoting healthier eating and drinking habits. \delete{The targets of these interventions reflect the key behaviors believed to be critical for promoting healthy eating and improving overall health outcomes.}\modif{Identifying the specific factors and behaviors targeted in prior research helps clarify health-relevant aspects of ingestion and associated intervention strategies. By analyzing these behaviors, we can evaluate how current interventions modify eating patterns, compare approach effectiveness, and uncover gaps in existing methods.}

This chapter aims to explore the targeted behaviors in current ingestive behavior intervention studies, \modif{and classify them} into contextual factors (external to the human body) and process behaviors (individual actions during the ingestion process). These two broad categories encompass a wide range of interventions designed to optimize various target factors or behaviors for ingestion and promote healthier ingestion habits.

\begin{figure}[htbp]
    \centering
    \includegraphics[width=\linewidth]{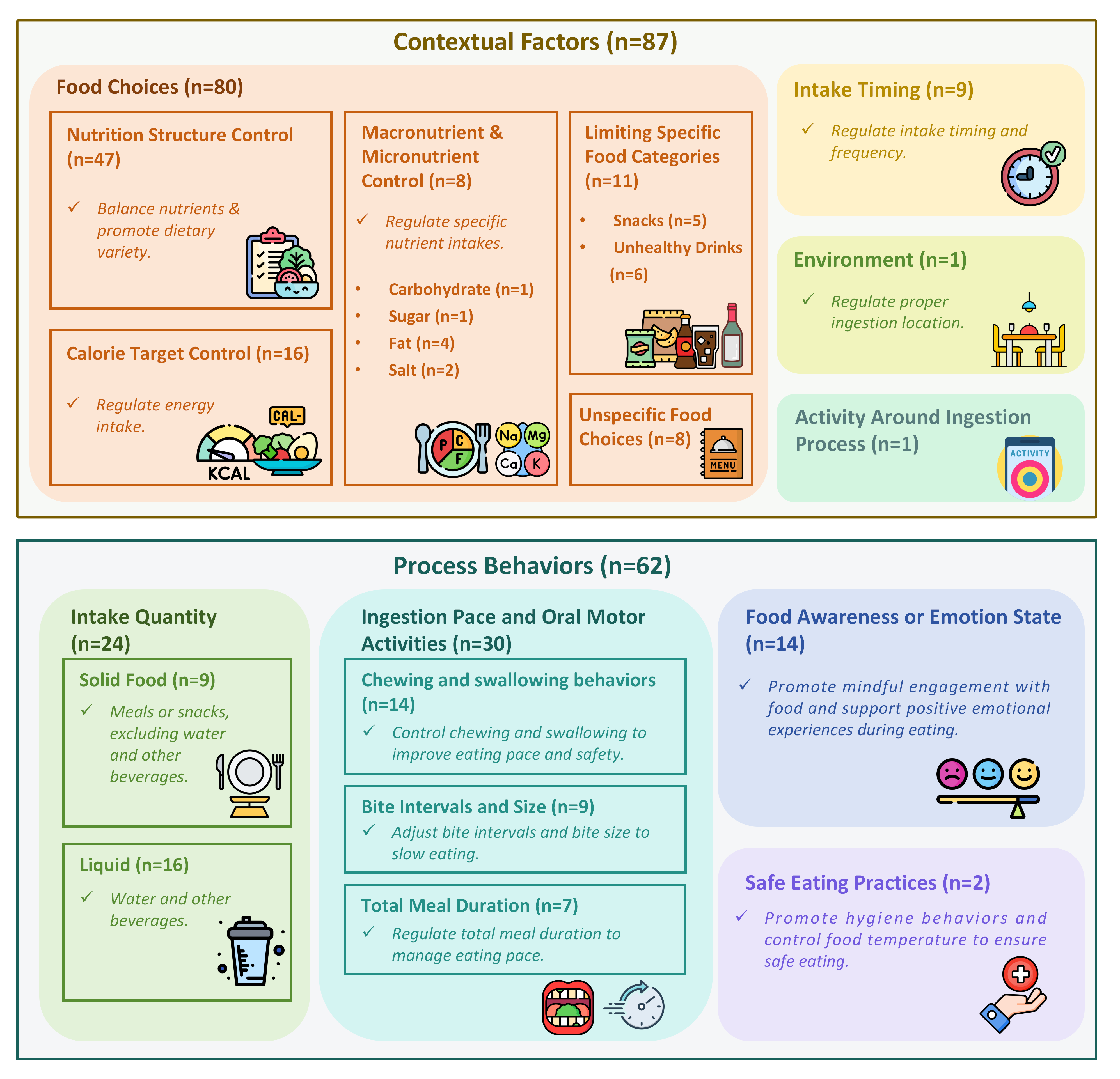}
    \caption{Taxonomy of target factors and behaviors with the number of studies involving each target.}
    \label{fig:target_taxonomy}
\end{figure}

\modif{Figure~\ref{fig:annual_count} presents the yearly distribution of intervention studies} which can be categorized into three groups: contextual factors only, process behaviors only, and studies targeting both. Both categories have garnered significant attention from researchers, with a greater number of interventions focusing on contextual factors. Building on this, Figure~\ref{fig:target_taxonomy} displays the taxonomy of these target factors and behaviors, providing a structured overview of the specific intervention targets within each category and the number of \modif{associated} studies. This taxonomy organizes the diverse intervention targets into a hierarchical structure, clarifying the relationship between broader categories and more specific target behaviors. A summary of these categories and the corresponding publications included in this review is shown in Table~\ref{tab:taxonomy}.

\subsection{Contextual Factors}

Contextual factors refer to external influences that affect ingestive behaviors, which are usually not limited to the duration of the ingestion process itself. These \delete{factors, which }include food choices, intake timing, the eating environment, and activities surrounding the ingestion process, \add{all of which} can shape an individual’s dietary patterns and habits. \delete{Since these influences often operate outside the immediate context of the meal, they play a crucial role in determining overall ingestive behaviors and are frequently targeted in interventions aimed at promoting healthier eating habits.}

\subsubsection{Food Choices}

Food choices play a fundamental role in shaping dietary patterns and are directly linked to the types and quality of nutrients consumed, significantly impacting long-term health outcomes. Poor food choices are strongly associated with the development of chronic diseases, such as obesity, diabetes, cardiovascular disease, and certain cancers, while healthier selections promote better metabolic health and disease prevention.

\delete{Given the central role of food choices in overall dietary health, many current ingestive behavior interventions prioritize food choices as a primary target.} In fact, \modif{interventions focusing on food choices represent the largest portion of the reviewed literature.} These interventions aim to improve dietary quality by modifying nutrient intake, promoting healthier eating habits, and restricting the consumption of unhealthy foods. Within these interventions, food choice-related targets are commonly categorized into several specific areas: Nutrition Structure Control\delete{(aiming for balanced nutrition and dietary variety)}, Calorie Control\delete{(regulating total energy intake)}, Macronutrient and Micronutrient Control\delete{(focusing on specific nutrients such as carbohydrates, fats, sugars, and salt)}, Limiting Specific Food Categories\delete{(e.g., reducing intake of snacks and sugary beverages)}, and Unspecified Food Choices\delete{(general recommendations for healthier eating without specific nutrient guidelines)}. Among these categories, interventions focusing on nutrition balance and calorie control are particularly prevalent, reflecting their perceived critical role in promoting healthy \modif{ingestion} and preventing diet-related chronic diseases.

\add{\textbf{i.}} \textbf{Nutrition Structure Control}: \modif{This target behavior aims to promote balanced nutrient intake by encouraging dietary diversity across major food groups, such as fruits, vegetables, whole grains, proteins, and dairy. A significant portion of interventions targets children, seeking to cultivate healthy eating habits early in life. Additionally, some studies address picky eating behaviors that limit dietary variety. By broadening food intake, these interventions help prevent nutrient deficiencies and support long-term health maintenance.}

\add{\textbf{ii.}} \textbf{Calorie Target Control}: 
\modif{This target behavior focuses on regulating energy intake through portion control, meal planning, and calorie management. }
\modif{Interventions guide individuals toward lower-calorie, nutrient-dense food, such as fruits, vegetables, and lean proteins, while encouraging the replacement of high-calorie items with healthier alternatives. }In the reviewed literature, calorie control is frequently employed as a strategy for weight management and \modif{obesity prevention}, underscoring its central role in ingestion health interventions.

\add{\textbf{iii.}} \textbf{Macronutrient and Micronutrient Control}: This category includes interventions targeting the regulation of specific nutrient intakes, such as carbohydrates, fats, sugars, and salt. Many interventions focus on adjusting these components to address health issues like diabetes, hypertension, and cardiovascular diseases. \delete{For example, regulating carbohydrate intake is often emphasized for individuals at risk of or diagnosed with diabetes, while reducing fat (particularly saturated fat) and sodium intake is associated with cardiovascular risk reduction.}

Current intervention targets include control over specific types of nutrients, such as:
\begin{itemize}
  \item \textit{Carbohydrate Intake} --- Regulating or reducing carbohydrate consumption can improve glycemic control and reduce the risk of chronic diseases, particularly among individuals with or at risk of diabetes \cite{reynolds2019carbohydrate}. Current interventions mainly target individuals with diabetes \cite{arefeen2024glyman}.
  \item \textit{Sugar Intake} --- Excessive sugar consumption is associated with obesity, diabetes, and dental problems \cite{moynihan2016sugars}. Current interventions associate sugar intake with glucose level \cite{liang2024exploring}.
  \item \textit{Fat Intake} --- Reducing total fat intake, especially saturated and trans fats, is essential for mitigating the risk of cardiovascular diseases and obesity, conditions strongly linked to excessive consumption of these unhealthy fats \cite{astrup2020saturated}. Current interventions often regulate fat intake to prevent obesity \cite{Vandelanotte2005EfficacyOS,Haerens2007TheEO,Burke2011TheEO,chukwu2011personalized}.
  \item \textit{Salt Intake} --- Limiting sodium intake helps prevent hypertension and cardiovascular risks \cite{jaques2021sodium}. Current interventions focus on controlling salt intake \delete{to promote healthier eating habits }in patients with metabolic syndrome, such as those with cardiovascular disease \cite{eyles2017salt, kim2016ecomeal}.
\end{itemize}

\add{\textbf{iv.}} \textbf{Limiting Specific Food Categories}: Some interventions specifically target the reduction or elimination of certain food categories considered unhealthy, such as snacks high in sugar or fat, and sugary beverages like soft drinks and alcohol. \delete{By limiting the intake of these foods, the aim is to reduce empty calorie consumption, lower the risk of obesity, and improve overall diet quality. This approach often complements broader nutrition and calorie-focused strategies by removing key sources of dietary excess.}

Current research focuses on reducing intake of:
\begin{itemize}
  \item \textit{Snacks} --- Interventions discourage high-calorie, low-nutrient snack consumption to reduce unnecessary caloric intake \cite{Bi2016AutoDietary:AW,liu2024hicclip,maramis2014preventing,bedri2022fitnibble,robinson2020social}.
  \item \textit{Unhealthy Drinks} --- Interventions mainly focus on alcohol intake \cite{matsui2018light,robinson2021humanoid} or sugar-sweetened soft drink intake control \cite{faltaous2021wisdom,Haerens2007TheEO,ritschel2018drink,fuchs2019impact}.
\end{itemize}

\add{\textbf{v.}} \textbf{Unspecific Food Choices}: This refers to general recommendations that do not focus on specific nutrients or food categories. A smaller subset of interventions addresses food choices in a more general way, providing \delete{broad }guidance to "eat healthier" without specifying particular nutrients or food categories. These interventions typically focus on encouraging awareness about healthy eating principles and promoting better food decision-making habits, sometimes through educational tools \cite{prasetyo2020foodbot, fernandez2013virtual}, reminders \cite{abisha2017embedded, luo2021foodscrap}, or behavioral nudges \cite{huang2018chatbot,terziouglu2023influencing,eigen2018meal}.

\subsubsection{Intake Timing}
Intake timing refers to the timing factors of ingestion, which significantly influence metabolic processes, health, and satiety. Studies have consistently shown that meal timing and frequency are closely linked to overall metabolic health, including insulin sensitivity \cite{nu14091719}. Unhealthy eating patterns, such as irregular meal timing or late-night eating, have been associated with a range of adverse health outcomes. Disruptions to the natural timing of meals can contribute to the development of chronic diseases, including obesity, cardiovascular disease, and diabetes \cite{yoshida2018association, rask2012tissue}. Additionally, meal frequency — how often a person eats throughout the day — has been found to influence metabolic health. Appropriate meal frequency may promote weight management and improve insulin sensitivity and blood glucose levels \cite{holmback2010high, carlson2007impact}. \delete{Along with maintaining regular meal times, consistent fluid intake is also essential for supporting bodily functions, such as temperature regulation, nutrient transportation, and waste removal.}

In current approaches, target intake timing is typically defined by the regular intervals between meals or water intake for appropriate eating frequency or to prevent dehydration \cite{xie2023chibo,Bi2016AutoDietary:AW,fortmann2014waterjewel,varun2023smart,lessel2016watercoaster,ravindran2022hydrationcheck}, as well as the optimal timing for meals consumption throughout the day to promote regular and healthy eating habits \cite{liang2024exploring,faltaous2021wisdom,ye2016assisting}.

\subsubsection{Environment}
\modif{Ingestive behaviors are shaped by a combination of social, cultural, and physical environments, which influence food choices, eating patterns, and the broader eating experience.} \delete{For example, the availability of healthy food options, social influences, and food cues in the surroundings can either promote or hinder healthy ingestive behaviors. }Research has shown that eating location and the surrounding environment can notably affect the intake of unhealthy foods. For instance, higher intakes of unhealthy snacks, such as those rich in sugars, fats, and sodium, are often observed in less structured eating environments like schools and while socializing \cite{casey2021added}.

Current interventions targeting the environment may address behaviors such as eating quickly while standing in the kitchen or not allocating enough time for meals at the dining table, both of which can negatively affect healthy eating. These environmental interventions aim to create supportive eating settings that encourage mindful eating and healthier food choices \cite{faltaous2021wisdom}.

\subsubsection{Activity Around Ingestion Process}
These interventions involve promoting healthy behaviors surrounding the ingestion process, such as encouraging individuals to engage in physical activity after meals \cite{alexander2017behavioral}.

\subsection{Process Behaviors}
Process behaviors refer to the specific actions individuals perform during ingestion. Interventions targeting process behaviors typically occur within the duration of an ingestion episode, although they may also span a longer ingestion period. \modif{These behaviors-including intake quantity, ingestion pace, oral motor patterns, food awareness, emotional state, and safe eating practices—directly influence food intake and digestion.}\delete{Argeting these aspects aims to improve eating habits, prevent overeating, promote mindful consumption, and reduce eating-related safety risks.}

\subsubsection{Intake Quantity}
The quantity of solid food and liquids consumed is crucial for managing energy balance, supporting overall health, and preventing weight-related issues. Adequate regulation of solid food intake helps control energy consumption and maintain a healthy weight, while sufficient liquid intake is essential to prevent dehydration, support physiological functions, and may also influence appetite and caloric intake. Interventions targeting intake quantity typically focus on two key aspects: controlling solid food portion sizes and managing liquid intake to promote healthier eating and drinking behaviors.

\add{\textbf{i.}} \textbf{Solid Food}: 
\delete{In this context,}Solid food refers to the ingestion of meals or snacks, excluding water and other beverages. \modif{These interventions aim to enhance awareness and control of food quantities to prevent overeating and reduce obesity risk.} They also support healthy ingestive behaviors, including the prevention of eating disorders \cite{rolls2002portion, livingstone2014portion}. Interventions targeting solid food portion control often measure intake quantity by assessing changes in food mass before and after a meal \cite{xie2023chibo,chaudhry2016evaluation,chen2022sspoon,Alshurafa2015RecognitionON,Scisco2011SlowingBR,maramis2014preventing,farooq2017reduction,moses2023investigating,li2018exploring}.

\add{\textbf{ii.}} \textbf{Liquid}: 
\delete{In this category, }Liquid intake refers to the consumption of water and beverages. \modif{Most interventions in this category emphasize maintaining adequate hydration to support physiological functions and promote healthier beverage choices. These strategies may also aid in energy balance and weight management \cite{popkin2010water, daniels2010impact}.} Interventions focusing on fluid intake typically measure water consumption by tracking changes in volume, such as the amount consumed between two time points or the total daily intake \cite{faltaous2021wisdom,Alshurafa2015RecognitionON,kaner2018grow,poddar2024aqua,kreutzer2015base,varun2023smart,bobin2018smart,lessel2016watercoaster,yildiz2019wwall,chiu2009playful,ko2007mug,pankajavalli2017hydration,wijanarko2019fuzzy,tommy2017interactive,ravindran2022hydrationcheck,zhou2021mosswater}. 

\subsubsection{Ingestion Pace and Oral Motor Activities}
\delete{Ingestion pace and oral motor behaviors play a crucial role in regulating energy intake and promoting healthy eating patterns. }Interventions targeting \modif{ingestion pace and oral motor behaviors} focus on modifying the speed and manner in which food is consumed to enhance satiety, prevent overeating, and support weight management. Faster eating rates have been consistently associated with higher energy intake, increased risk of overweight and obesity, and poorer appetite regulation. A growing body of evidence supports the effectiveness of slowing down ingestion pace as a practical strategy for preventing obesity and related chronic conditions \cite{robinson2014systematic,slyper2021oral}. Current approaches typically target three main aspects: chewing and swallowing behaviors, bite intervals and size, and total meal duration. 

\add{\textbf{i.}} \textbf{Chewing and swallowing behaviors}: 
Chewing and swallowing behaviors are key components influencing the pace of eating and the overall ingestion process. \modif{Many interventions encourage individuals to chew more times per bite or slow down the rate of chewing. These practices can naturally slow their eating pace, enhance satiety signaling, and reduce overall food intake, which supports weight management and prevention of overeating} \cite{kadomura2013educatableware,kleinberger2023auditory,koizumi2011chewing,Bi2016AutoDietary:AW,Scisco2011SlowingBR,maramis2014preventing,chen2022slnom,sugita2018diet,turner2017byte,hori2023masticatory}. 
\modif{Others} combine the regulation of chewing and swallowing, aiming to control the full oral processing cycle rather than focusing solely on chewing. These approaches aim to slow ingestion by coordinating the timing between chewing and swallowing, ensuring thorough oral processing before taking the next bite \cite{kim2016slowee, kamachi2023eating}. 
Additionally, a few interventions focus on modifying oral motor behaviors by adjusting chewing force and modulating the swallowing phase. These interventions promote safer, more mindful eating practices, improving digestion and reducing the risk of eating-related complications \cite{biyani2019intraoral, zheng2025alteration}.

\add{\textbf{ii.}} \textbf{Bite Intervals and Size}: 
\modif{As critical components of eating pace regulation, bite size and inter-bite intervals have been intervened} to allow more time for satiety signals to develop, helping individuals recognize fullness earlier and thus reducing overall food intake. Most interventions focus on extending the time interval between successive bites, encouraging individuals to pause longer before taking the next bite \cite{nakamura2023eat2pic,de2020multimodal,lee2019user,kim2016eating,kim2018smartwatch,kim2018animated,mendi2013food,hermans2017effect}. \modif{Bite size is another key behavioral target. Reducing the amount of food per bite naturally slows eating pace and lowers overall caloric intake during a meal \cite{chen2022sspoon}.}

\add{\textbf{iii.}} \textbf{Total Meal Duration}: 
\modif{The overall time taken to complete a meal is typically estimated by tracking changes in food weight over time to provide an objective measure of ingestion pace.} Many interventions focus on prolonging meal duration to slow down eating speed, helping individuals better recognize satiety cues and prevent overeating\delete{, which supports healthier dietary behaviors} \cite{moses2023investigating, zhang2019applying, kim2016ecomeal, ioakimidis2009method}.
\modif{Others, particularly those involving young children, seek to avoid excessively long meals to help establish structured eating routines early in life \cite{krishna2025nutrifit, lo2007playful}.} 
Additionally, some interventions are designed to synchronize eating paces in social settings, where regulating total meal duration across a group helps reduce discomfort caused by uneven eating speeds \cite{mitchell2015really}.

\subsubsection{Food Awareness or Emotional State}
Interventions targeting food awareness or emotional state focus on enhancing individuals’ cognitive and emotional engagement with their eating experiences to promote healthier behaviors. A key approach \delete{in this area }is encouraging greater attention to food during meals, either through practices like mindful eating, which emphasizes being fully present during eating \cite{kleinberger2023auditory, koizumi2011chewing, parra2023enhancing, wang2025reflecting, khot2020swan}, or through interventions designed to help children develop better focus on their food \cite{kadomura2013educatableware, han2017childish}. 
\modif{Other interventions} focus on improving individuals’ self-perception and cognitive evaluation of their ingestive behaviors after meals. By increasing reflective thinking about eating habits, these interventions help individuals build healthier long-term dietary patterns and strengthen their self-regulation capacity \cite{sun2020postcard, zhang2020eat4thought, tang2015introducing}. 
Moreover, \modif{a separate} line of interventions aims to enhance the emotional quality of the eating experience itself, such as promoting positive emotions through novel experiences and reducing negative emotions like anxiety during eating. A healthy emotional state during ingestion has been linked to better digestion, improved dietary choices, and a lower risk of disordered ingestive behaviors \cite{mayumi2023design, carroll2013food, mitchell2015really, narumi2011augmented}.

\subsubsection{Safe Eating Practices}
Safe eating practices \modif{are intended to minimize health risks during the ingestion process.} Current interventions \delete{in this area }primarily target two aspects: hygiene behaviors and food temperature control. Hygiene-related interventions focus on promoting safe practices during eating to prevent foodborne infections \cite{wang2021research}. Other interventions address food temperature control to prevent oral injuries, such as burns \cite{singhal2024multifunctional}.



\section{Structures and Design Space \modif{from} Behavioral Sensing \modif{to} Intervention Paradigms}
\label{chapter:rq2}

\modif{This section analyzes the modality-level characteristics of ingestion health systems, focusing on two essential stages of the behavioral loop: sensing and intervention. While reasoning processes mediate between perception and action, they are typically embedded within pipelines, reflected in how sensing data is interpreted and applied. Therefore, our analysis centers on the sensing and intervention components, where the design of modality plays a direct and observable role in shaping user experience and effectiveness of the system.}

\modif{Prior research has established several perspectives on modality. \citet{jaimes2007multimodal} define modality as "\textit{modes of communication according to human senses and computer input devices activated by humans or measuring human qualities}", emphasizing the centrality of perception and input-output channels. Building on this view, \citet{jansen2022design} extend the concept by introducing cerebral and cardiac modalities to accommodate emerging interactive technologies. However, in ingestion-related systems, certain modalities do not correspond directly to individual human sensory channels but instead involve environmental interactions or indirect physiological monitoring. To address this, we introduce the category of \textit{environment-based modalities}, which encompass input and feedback mechanisms targeting environmental objects independently of human senses. This extension enables a more comprehensive classification of sensing and intervention designs in ingestion health interventions.}

\modif{Building on these foundations, we classify human-based modalities into \textbf{visual, auditory, haptic, gustatory, olfactory, motion}, and \textbf{physiological}, and environment-based modalities into \textbf{Visual, Acoustic, Force, Spectral}, and \textbf{Electrical}.} Slight structural and descriptive variations in modality classifications are introduced \modif{below} to align precisely with each phase of the behavioral closed-loop.

\subsection{Sensing}

This subsection focuses on sensing modalities within overall ingestion health intervention paradigms, where behavioral data are primarily acquired through various sensor-based or perceptual computing-driven modalities. The detailed structure and descriptions of each modality, along with its associated submodalities in the sensing process, are summarized in Table \ref{tab:sensing_modality_table}.

Notably, gustatory and olfactory modalities are absent in the sensing stage across existing systems. We believe that sensing gustatory or olfactory information through perceptual or sensor-based means remains technically challenging, and these modalities are mainly employed in the \delete{feedback/}intervention phase rather than for direct behavioral sensing. \modif{The identified sensing modalities is categorized according to their classifications below.} \add{A summary of sensing modalities and corresponding publications included in this review is shown in Table \ref{tab:sensing modality paper}.}

\subsubsection{Human-based}
\

\textbf{\add{i.} Visual}: Visual sensing has been widely adopted in ingestion-related systems to extract behavioral and attentional cues through camera-based observation. We identified three common submodalities: \delete{within visual sensing: gaze, facial features, and body movement.}

Gaze\delete{-based} sensing tracks eye movements and fixation points to assess users’ attention during food-related interactions. \modif{It has been used to measure attentional engagement with food stimuli across settings such as VR eye tracking \cite{karkar2018virtual} or webcam-based monitoring \cite{khot2020swan}.}


\modif{Facial-expression sensing primarily support both emotional and behavioral inference in ingestion-related systems. It has been used to detect affective states \cite{mayumi2023design}, model chewing through lip and jaw motion \cite{kim2018animated}, and recognize user presence in multi-user contexts \cite{triantafyllidis2019social, robinson2020social}.}


\modif{Body posture sensing interprets behaviors by capturing body and head orientation, alignment, and movement during ingestion. Posture-related signals were captured by tracking the feeding gestures of children for satiety signals \cite{xie2023chibo}, monitoring the upper-body orientation using skeleton models \cite{mayumi2023design}, and analyzing egocentric video for mealtime behavior \cite{zhang2020eat4thought}.}





\textbf{\add{ii.} Auditory}: Auditory sensing in ingestion health systems typically \modif{captures} either verbal input or eating-related sounds, which can be classified into two main submodalities:\delete{ speech and non-verbal sound, covering existing approaches.}

Speech\delete{-based auditory} sensing \modif{commonly supports} semantic data acquisition and\delete{ enabling} conversational interfaces. \modif{Spoken input has been used to extract structured dietary data via linguistic pattern recognition \cite{prasetyo2020foodbot}, facilitate food-related recommendations from spoken queries \cite{mayumi2023design, robinson2020social}, and log intake events with interaction \cite{luo2021foodscrap, arefeen2022computational}, as well as to enable system-guided prompting via Wizard-of-Oz protocols \cite{parra2023enhancing}.}


\modif{Non-verbal sound sensing focuses on physiological cues like chewing or swallowing. Microphones have been placed near the throat \cite{kalantarian2014spectrogram} or collar \cite{chen2022slnom} to detect bites \cite{Bi2016AutoDietary:AW}, classify food types \cite{kleinberger2023auditory}, or estimate intake rhythms. Earpiece-mounted sensors, including chewing microphones \cite{maramis2014preventing} and bone conduction devices \cite{kamachi2023eating}, help reduce ambient noise. Some systems further adapt to context, such as denture-based chewing detection \cite{koizumi2011chewing} or setups capturing both speech and eating sounds \cite{zhang2020eat4thought, kamachi2023eating}.}





\textbf{\add{iii.} Haptic}: Haptic sensing aims to detect physical interactions or internal mechanical activity through pressure, force, or contact-based sensors. \modif{A common application is weight sensing, which serves as a long-term indicator of dietary intake \cite{ludwig2011refresh, bastida2023promoting, nedungadi2018personalized}. Wearable piezoelectric and force sensors have been used to capture throat vibrations during swallowing \cite{Alshurafa2015RecognitionON, kim2016slowee, abisha2017embedded} and temporalis muscle displacement for chewing estimation \cite{farooq2017reduction}. Other methods include intraoral pressure sensors embedded in palatal retainers to measure tongue force \cite{biyani2019intraoral}, external sensors for estimating masticatory force \cite{zheng2025alteration}, and pressure-sensitive interfaces for logging ingestion events \cite{fortmann2014waterjewel, tang2015introducing, cai2024see, robinson2020social}.}

\textbf{\add{iv.} Motion}: Motion sensing captures gestures and subtle physical activity using inertial or proximity sensors.

Most gesture-oriented systems employ inertial measurement units (IMUs) mounted on \modif{utensils \cite{wang2025reflecting, khot2020swan, zhang2019applying, li2018exploring, kadomura2013sensing, kadomura2014persuasive, hermans2017effect}, containers \cite{tommy2017interactive, bobin2018smart, chiu2009playful, zhou2021mosswater}, wristbands \cite{kim2016eating, kim2018smartwatch, ye2016assisting, turner2017byte, mendi2013food, kumbhare2023low, gaikwad2024precision, dragoni2017semantic, Scisco2011SlowingBR}, smart glasses \cite{bedri2022fitnibble} and smartphones \cite{tirasirichai2018bloom} to detect hand-to-mouth movement, bite frequency, or drinking actions.} Some platforms also use pedometers \cite{harous2017hybrid} and activity trackers \cite{alexander2017behavioral, vazquez2015development, fuchs2019impact} 
to estimate ingestion or physical activity context. These approaches have made it easier for broader health-monitoring systems to achieve dietary reasoning support.

Chewing and micro-head motion sensing target jaw, throat, or mouth-level interactions. \modif{\citet{biyani2019intraoral} tracked throat motion using a 3D accelerometer, while \citet{hori2023masticatory} used infrared sensors to detect subtle skin vibrations. Sensors based on electrical conductivity \cite{kadomura2013sensing, kadomura2013educatableware} or static charge \cite{han2017childish} have also been embedded in utensils and cups to infer mouth contact and drinking events.}

\textbf{\add{v.} Physiological}: Physiological sensing provides insight into users’ internal states that influence or contextualize ingestion behavior, including appetite, metabolic status, stress, and cardiovascular conditions. 

Autonomic sensing typically involves EMG electrodes to detect muscle activity during chewing or wrist motion \cite{lee2019user, sugita2018diet, kim2016slowee}, \modif{along with} thermistors and respiratory sensors to track body temperature and breathing patterns \cite{park2007development, jung2007wellbeing}. Emotional arousal or stress is inferred through electrodermal activity (EDA) sensors \cite{carroll2013food, nedungadi2018personalized}\delete{, which may affect eating pace or impulsive behaviors}.

Metabolic sensing employs BIA sensors to estimate body fat composition  \cite{chukwu2011personalized, balbin2020scientific}, and incorporates CGM and cholesterol sensors to continuously monitor glucose levels and lipid markers  \cite{liang2024exploring, harous2017hybrid, arefeen2024glyman, nedungadi2018personalized}. 

Cardiac sensing captures heart rate, blood pressure, and oxygen saturation using ECG  \cite{badawi2012real}, pulse wave  \cite{varun2023smart}, or optical PPG sensors  \cite{park2007development, ludwig2011refresh, jung2007wellbeing, huang2018chatbot, nedungadi2018personalized} \modif{to support context-aware interpretation and intervention timing.}





\textbf{\add{vi.} Others}: \modif{Without sensing or perceptual computing, other approaches rely entirely on manually collected self-reports.} \modif{Common inputs include manual logging of} dietary events or physical states, image selection or annotation, and questionnaire-style \delete{logging }via mobile or chatbot interfaces. \modif{Though such} systems \modif{can} \modif{support reasoning and intervention,} they fall outside our \modif{sensing stages} due to the lack of automated perceptual input or computational sensing logic. \delete{A detailed overview of this manual approach is summarized in Table \ref{tab:sensing modality paper}, with approximately 20 publications adopting this method.}




\subsubsection{Environment-based}
\

\textbf{\add{i.} Visual}: Environment-based visual sensing focuses on object-level perception of food and eating-related artifacts to support dietary monitoring and context acquisition. Most systems \modif{apply} computer vision to classify food types, estimate portions sizes, or infer \modif{nutrition} from images \cite{kanjalkar2024ai, eigen2018meal, cordeiro2015rethinking, yr2024nutrispy, epstein2016crumbs}\modif{ ,via} smartphone photography \cite{mishra2024hybrid, waltner2015mango, achilleos2017health}, utensil- or refrigerator-mounted cameras \cite{nakamura2023eat2pic, kwon2016smart, faltaous2021wisdom}, or wearable modules \cite{fuchs2019impact}.


\modif{Other methods use food color \cite{kadomura2013sensing, kadomura2014persuasive}, utensil-based cues \cite{wang2021research}, visual distortion \cite{akter2021foodbyte, chiu2009playful}, or plate dynamics \cite{ganesh2014foodworks} to infer consumption. Additional approaches include receipt parsing \cite{mankoff2002using}, puzzle-based interaction \cite{cai2024see}, packaging input \cite{arefeen2022computational, hsiao2011intelligent}, and visual code recognition such as barcode scanning \cite{volkova2016smart, eyles2017salt}, QR code interpretation \cite{fernandez2013virtual, vazquez2015development, alloghani2016mobile}, and edible marker detection in AR scenarios \cite{gutierrez2018phara}.}

\textbf{\add{ii.} Acoustic}: Acoustic sensing often utilizes ultrasonic distance sensors to monitor liquid intake or food-related actions. 
\modif{Most systems embed ultrasonic transducers in bottles or containers to track fluid levels and estimate drinking volume \cite{poddar2024aqua, wijanarko2019fuzzy, ravindran2022hydrationcheck, varun2023smart, tommy2017interactive}. Others apply similar ultrasonic sensors to detect discrete events like snack removal \cite{liu2024hicclip}, or monitoring food depletion in storage cabinets \cite{faltaous2021wisdom}.}


\textbf{\add{iii.} Force}: Force sensing typically \modif{uses} load cells or pressure sensors embedded \modif{to} measure changes in food or liquid weight over time. Smart \modif{utensils} equipped with load cells \cite{mitchell2015really, kim2016ecomeal, maramis2014preventing, ioakimidis2009method, de2020multimodal, lessel2016watercoaster, ritschel2018drink} or force-sensitive resistors (FSRs) \cite{jung2017mom} have been used to continuously track food \add{or liquid} depletion and infer \modif{ingestion} pace or satiety. \modif{Building on ingestion tracking, systems aim to estimate nutrient intake \cite{krishna2025nutrifit, zhao2021funeat, de2020multimodal, jung2017mom}, detect food preferences and portion control \cite{joi2016interactive, hermans2017effect}, and even gamify eating through weight-sensitive interactions \cite{lo2007playful}.}


\textbf{\add{iv.} Spectral}: Spectral sensing \modif{commonly leverages} infrared detectors, temperature sensors, and light-spectrum analyzers, to capture physical or chemical properties of food, beverages, and environmental conditions.
\delete{ Specifically,} \modif{Infrared} or optical sensors have been used to detect the presence of containers \cite{ritschel2018drink}, estimate proximity to food \cite{bedri2022fitnibble}, or assess food temperature  \cite{singhal2024multifunctional}. Temperature sensors embedded in \delete{bottles or }utensils help monitor liquid warmth \cite{tommy2017interactive, wijanarko2019fuzzy, varun2023smart} or environmental conditions affecting food stability  \cite{bastida2023promoting, badawi2012real, kwon2016smart}. \modif{In addition, \citet{matsui2018light} used spectral absorption to estimate alcohol content and drinking rate.}


\textbf{\add{v.} Electrical}: 
Electrical sensing supports ingestion\delete{-related} monitoring by detecting changes in conductivity, capacitance, or magnetic response. \modif{It enables passive contact-based tracking without user instrumentation.} Common sensor\add{s} \delete{types} include conductivity probes, humidity sensors, capacitive pads, and RFID/NFC modules.

\modif{Conductivity and humidity sensors embedded in utensils can} monitor water levels and drinking events \cite{bobin2018smart, tommy2017interactive, kreutzer2015base, varun2023smart}, while capacitive \modif{sensing can estimate volume via dielectric change \cite{kaner2018grow}.} Some systems integrate float switches or conductive tilt detection \modif{to identify bottle usage \cite{pankajavalli2017hydration, ko2007mug}.} Ambient humidity \modif{has also been} used to characterize dining environments and inform food safety assessments \cite{bastida2023promoting, badawi2012real, kwon2016smart}.

\modif{Magnetic technologies} enable automatic identification of food items, storage states, or utensil presence  \cite{vazquez2015development, huang2010context, jung2017mom}. Some advanced \modif{RFID or NFC designs} estimate consumed volume via passive milliliter tracking \cite{yildiz2019wwall}. Additionally, \citet{han2017childish} used metal contact detection to support real-time eating activity sensing.



\subsection{\delete{Feedback and }Intervention}
\delete{As the final stage in the ingestion behavior loop, feedback and }Intervention mechanisms\add{, whether active or passive,} convey information back to users to support behavioral awareness, regulation, or change. \delete{In contrast to sensing, which focuses on data acquisition through sensors or perceptual computing, intervention emphasizes human interpretation and experience.} To systematically present the design of the \delete{feedback and }intervention process, we do not assess whether these mechanisms rely on dedicated sensors. Rather, since all systems included in our review engage in sensor-based or perception-driven processing at some point in the ingestion-related loop, we categorize their \delete{feedback and }intervention strategies in terms of modality, focusing on how information is conveyed and perceived. \add{To provide an overview of the types of modalities employed, Table \ref{tab:feedback/intervention modality paper} summarizes a taxonomy of intervention modalities and their corresponding publications.} \modif{Accordingly, the modalities at this stage are inherently based on the primary human sense they target to deliver prompts, guidance, or other affective cues,} as summarized in Table \ref{tab:feedback_intervention_modality_table}.

\subsubsection{Visual}
Visual \modif{cues} is one of the most commonly used modalities in ingestive behavior interventions, providing intuitive, accessible, and scalable means of conveying prompts, progress, or state awareness. Based on representation format, we categorize visual \modif{cues} into three submodalities:\delete{ textual, graphical, and symbolic. This layered approach to visual design allows systems to balance clarity, immediacy, and ambient subtlety, tailoring feedback complexity to user needs and situational demands. A complete list of visual feedback systems and their classified modalities is available in Table \ref{tab:feedback/intervention modality paper}.}

\textbf{\add{i.}Textual}:
\modif{Text-based systems convey dietary information via notifications, reminders, or instructions. They often support user behavior by prompting actions, tracking progress, and aiding decisions. Examples include recommender agents \cite{zenun2017online, arefeen2022computational}, JIT assistants \cite{prasetyo2020foodbot, carroll2013food}, and self-monitoring apps \cite{Burke2011TheEO, krishna2025nutrifit}. Many others use dashboards or push notifications to present progress and reinforce goals \cite{faltaous2021wisdom, lee2019user, Haerens2007TheEO}.}

\textbf{\add{ii.}Graphical}:
\modif{Graphical methods use visual elements, such as charts, bars, or data views, to convey behavioral patterns and intervention outcomes. These representations support self-awareness by illustrating nutritional balance, eating pace, or meal composition. For example, eat2pic transforms bite patterns into visual art \cite{nakamura2023eat2pic}, while others visualize ingestion summaries through app dashboards \cite{faltaous2021wisdom, Scisco2011SlowingBR, lo2007playful}.}

\textbf{\add{iii.}Symbolic}:
Symbolic approaches convey prompts \modif{with} simplified visual representations, such as color changes \cite{bobin2018smart}, flashing lights \cite{matsui2018light, lessel2016watercoaster}, or emoji-style icons \cite{kim2018animated}, which communicate behavioral cues with minimal cognitive load. These signals are often embedded in utensils or ambient displays to indicate states like "ready to eat" or "slow down\delete{ without requiring linguistic interpretation} \cite{agarwal2024user, kim2016slowee}.

\subsubsection{Auditory}

Auditory modalities primarily rely on simple sound output devices, most commonly speakers embedded in utensils, tableware, or ambient systems. \modif{These systems deliver symbolic or affective auditory cues that shape perception, pacing, and awareness during meals. Such cues influence eating through cross-modal effects --- for example, enhancing chewing sounds can help slow intake \cite{kleinberger2023auditory}. Interactive forks and tableware produce feedback tones or playful audio to encourage mindful eating, particularly in children \cite{kadomura2013educatableware, cai2024see}. Auditory prompts are also employed in game-based or robot-assisted contexts to guide intake timing \cite{xie2023chibo, kadomura2014persuasive}. This modality provides a lightweight, ambient pathway that can be unobtrusively integrated into multisensory interventions.}

\subsubsection{Haptic}
Haptic modality involves bodily interventions using physical signals such as vibration or \delete{shape }deformation. \modif{Rather than} relying solely on visual or auditory channels, \modif{it} engages \delete{the user’s }sensorimotor experience by delivering tactile cues, modifying utensil mechanics, or altering physical resistance to shape real-time food-related actions.

\textbf{\add{i.}Vibration}: 
\modif{Most systems use vibration motors or piezoelectric actuators embedded in utensils, wearables, or phones, to offer a non-disruptive alternative to visual or auditory cues. Utensil-based systems include bottles that vibrate to prompt hydration \cite{faltaous2021wisdom, varun2023smart}, chopsticks for posture correction \cite{wang2021research}, and cutlery for eating speed regulation \cite{hermans2017effect}. Wearables like wristbands or smartwatches deliver vibrotactile intervention for fast eating or poor chewing \cite{kim2016slowee, kim2016eating, kim2018smartwatch, kamachi2023eating}. Some systems also use phone vibrations to prompt hydration \cite{chiu2009playful} or journaling \cite{ye2016assisting}.}


\textbf{\add{ii.}Deformation}: 
\modif{Deformative methods typically modify physical properties of devices via pneumatic or motorized actuation to influence ingestive behavior. Utensil designs like SSpoon \cite{chen2022sspoon}, SWAN \cite{khot2020swan}, and other deformable spoons \cite{zhang2019applying, singhal2024multifunctional} adjust posture or resistance to slow intake or enhance control. Wearables such as FatBelt simulate abdominal fullness through inflation after caloric thresholds \cite{pels2014fatbelt}. Other systems use deformable materials, such as VR rubber donuts \cite{li2018exploring} or scent-releasing tactile surfaces \cite{cai2024see}, to influence ingestive behavior.}


\subsubsection{Olfactory}

\modif{Olfactory modality uses scent stimuli to influence perception, appetite, or food-related associations, typically through intentional odor delivery. Methods range from digital aroma delivery to embedded scent materials, including VR-synced diffusers to curb intake \cite{li2018exploring}, odor filters to shift food perception \cite{narumi2011augmented}, and scent puzzles to aid food recognition in children \cite{cai2024see}.} \modif{These} approaches highlight the affective and cognitive potential of olfactory \modif{cues}, especially when combined with visual or tactile input.


\subsubsection{Physiological}


Among physiological modalities, electrodermal stimulation is the only technique used in ingestion interventions, \modif{delivering intraoral electrical pulses to modulate muscle activity. \citet{zheng2025alteration} stimulated the mandibular vestibule based on chewing force, while \citet{biyani2019intraoral} triggered soft palate stimulation via tongue pressure to enhance swallowing. Though rarely explored, such sensorimotor methods target neuromuscular control and show promise for low-latency modulation of ingestion behavior.}


\subsection{Design Space Mapping through Modality to Behavior}
\label{chapter:rq2_3}

To further characterize the distribution and coverage of closed-loop \modif{paradigms} for sensing and intervention in ingestion health, we conducted a morphological analysis combining the dimensions of sensing and \delete{feedback/}intervention modalities with their corresponding target behaviors, inspired by  \citet{zwicky1967morphological} and  \citet{jansen2022design}.

This results in two-dimensional design matrices, shown in Table \ref{tab:sensing-behavior} and Table \ref{tab:intervention-behavior}, \modif{mapping each modality (x-axis) to specific behavioral targets (y-axis).} To reduce redundancy and \modif{improve clarity}, we adopt a mixed-level classification along the modality axis, \modif{collapsing fine-grained subcategories with overlapping implementations into higher-level groupings reflecting functional similarity.} The matrices enable us to identify well-covered modality–behavior pairs, underexplored combinations, and \delete{potential} design opportunities. \modif{This visualization can serve as a framework to inspire new behavior-aware interventions in ingestion health.}

\textbf{Analysis of Table \ref{tab:sensing-behavior}} reveals \delete{several key }trends in \delete{the }current \modif{sensing designs}. Human-based motion sensing, \modif{especially} wrist and object trajectory tracking, demonstrates broad behavioral coverage \modif{of} oral and finger motor activities \delete{and some other context factors}. \modif{Visual and acoustic modalities are also common for food recognition, intake timing, and pacing, often using image, sound, or AR inputs.} \modif{In} contrast, environment-based modalities like electrical or magnetic sensing remain underutilized\delete{, appearing in only a few cases}. Physiological sensing (e.g., EMG, EDA, CGM) \modif{provides} rich internal state signals but is \modif{mainly} limited to \modif{emotional or metabolic monitoring, suggesting potential for broader integration.} The matrix also highlights underexplored areas, including food hygiene, emotional triggers, and environmental dynamics. These gaps point to opportunities for multimodal fusion and novel device placements. Overall, the mapping reveals both saturated modality–behavior pairs and areas where new sensing strategies can enhance behavior-aware intervention design.

\textbf{Analysis of Table~\ref{tab:intervention-behavior}} reveals trends in \modif{intervention} modality distribution. Textual and graphical visual \modif{intervention} dominate, especially for food choices, intake quantity, and ingestion pace\modif{, favored for their accessibility and clarity.} Symbolic \modif{intervention} appears less frequently but with \modif{increasing} diversity. LED indicators are used across hydration, eating pace, and safety contexts. Symbolic auditory \modif{cues, }such as chewing sounds or metaphorical \modif{signals}, \modif{serve} as a lightweight and engaging method \modif{to promote} awareness. Tactile
symbolic forms, like thermochromic or pneumatic responses, remain niche but offer ambient and intuitive signaling. Auditory \modif{interventions} via speakers and alarms spans nearly all behavior types, while haptic \modif{interventions}, \modif{mainly} motor-based, \modif{focus on} emotional regulation and pace control. More advanced haptics (e.g., piezoelectric, pneumatic) remain \modif{underused}. Experimental modalities like olfactory and physiological \modif{interventions} appear in a few behaviorally rich cases, offering unique multisensory pathways. Overall, the table highlights the dominance of visual modalities but also \modif{suggests} symbolic and sensory alternatives as promising design directions. Gaps in environmental, emotional, and hygiene-related \modif{cues} \modif{indicate} potential for broader and more creative \modif{intervention} strategies in closed-loop ingestion health systems.

Notably, despite the variety of sensing pathways, few systems reuse the same sensors for both perception and intervention. Most existing \modif{intervention} channels \modif{still} rely on smartphones or static screens rather than leveraging interactive or perceptual sensors for embodied or context-aware interventions. This reflects a substantial unexploited design space, which we will discuss \modif{further in Chapter~\ref{chapter:discussion}, including specific hypotheses about effective sensing–intervention combinations generated from the design space.}

\section{Effectiveness and User Acceptance of Current Ingestion Health Intervention}
\label{effectiveness}

In this section, we examine papers that include user studies to assess the effectiveness and user acceptance of their work. To systematically understand their results, we have categorized these papers into four types: \textit{behavioral \delete{feedback/}intervention}, \textit{educational \delete{feedback/}intervention}, \textit{social \delete{feedback/}intervention}, and \textit{other specific \delete{feedback/}intervention}. Furthermore, we have divided each type into three paradigms: \modif{passive intervention, active intervention, and combined intervention which includes both passive and active intervention.} In total, 106 papers are included.

\subsection{Behavioral \delete{Feedback/}Intervention}

Behavioral \delete{feedback/}intervention refers to approaches that primarily focus on altering specific ingestive behaviors of users, such as eating speed, chewing counts, and drinking frequency\delete{, ultimately assisting them in achieving healthier eating habits}. The technical approaches in this category primarily rely on sensor-based technologies \delete{(e.g., IMU, piezoelectric sensors, RFID) }and real-time feedback mechanisms\delete{(e.g., visual, haptic, and auditory cues)}.


\subsubsection{\modif{Passive Behavioral Intervention}}

\modif{Passive behavioral intervention presents real-time or retrospective data reflecting a user's actions, fostering self-awareness and autonomous behavior change without direct instruction.} Systems that provide \modif{passive behavioral intervention} --- such as continuous glucose monitoring \cite{liang2024exploring} and smart hydration bracelets \cite{fortmann2014waterjewel} --- offer unobtrusive data visualization without explicit instructions\delete{, relying on users' self-reflection to drive behavioral changes}. \modif{Also}, ambient logging devices \cite{tang2015introducing} and smart nutrition trackers \cite{Alshurafa2015RecognitionON} collect and display consumption patterns through minimal interfaces, facilitating autonomous behavior adjustment. A total of 20 papers are included in this category.

Feedback effectiveness is evidenced by measurable behavioral changes. For instance, real-time glucose feedback \cite{liang2024exploring} reduced average glucose levels in 50\% of participants and prompted adjustments in meal timing\delete{, aligning with the "action" stage of the transtheoretical model}. Bite-rate feedback \cite{kim2016slowee, Scisco2011SlowingBR} led to a 57.7\% reduction in energy intake among fast eaters\delete{, although the effects varied based on baseline consumption habits}. Isomorphic feedback, such as the waist inflation from FatBelt \cite{pels2014fatbelt} and ambient displays \cite{kaner2018grow}, utilized physical or visual cues to reduce calorie intake and increase water consumption by 20–30\% respectively. Notably, the modality of \modif{passive intervention} matters: younger users \delete{(ages 20s) }preferred vibration feedback, while older adults \delete{(ages 50–70) }favored light-based feedback \cite{kim2016slowee}. Additionally, manual input requirements, such as the time-interval logging in the SAL logger \cite{tang2015introducing}, can reduce adherence, particularly among older adults.

\modif{Passive behavioral intervention} demonstrates consistent patterns in user acceptance\delete{ and effectiveness} across multiple studies. \modif{Passive behavioral intervention succeeds by balancing autonomy and personalization, with its seamless daily integration outperforming those requiring active engagement.} For instance, wrist-worn devices, such as the WaterJewel bracelet \cite{fortmann2014waterjewel} and the IDEA wristband \cite{lee2019user}, achieve a long-term usage willingness of 70–83\% due to their non-intrusive design and plug-and-play functionality. In contrast, neck-worn \cite{fortmann2014waterjewel} or head-mounted sensors face lower acceptance rates due to discomfort or social stigma \cite{lee2019user}. Additionally, aesthetic customization, such as WaterJewel’s design catering to both male and female users \cite{fortmann2014waterjewel} and the GROW bottle’s thermo-chromic display \cite{kaner2018grow}, further enhances acceptance by aligning with personal preferences.

\subsubsection{\modif{Active Behavioral Intervention}}

\modif{Active behavioral intervention} actively modifies ingestive behaviors through direct mechanisms, such as physical constraints or real-time prompts. Physical constraint systems, like the pneumatic shape-changing spoon \cite{chen2022sspoon}, mechanically regulate portion sizes and chewing rates by modifying the tangible interface. Digital guidance platforms, such as portion estimation interfaces \cite{chaudhry2016evaluation}, provide prescriptive dietary recommendations via mobile apps and interactive displays. More immersive interventions, including the sensory interactive table \cite{de2020multimodal} and the smart tray system \cite{jung2017mom}, combine environmental sensing with real-time behavioral prompts to actively reshape dining interactions. \delete{Unlike passive feedback that merely presents data, these interventions enforce immediate behavioral changes by altering how users interact with food. }A total of 15 papers are included in this category.

\modif{Active behavioral intervention} systems actively reshape user actions through physical, computational, or gamified means. Interventions that physically alter user behavior demonstrate particularly strong efficacy. For instance, the Spoon's shape-changing mechanism reduced bite size by 13.7–16.1\% and total intake by 4.4–4.6\% without affecting appetite \cite{chen2022sspoon}. Additionally, intraoral closed-loop systems increased mastication force by 26\% through real-time electrical stimulation \cite{biyani2019intraoral} and improved swallowing duration via neuromuscular triggers \cite{zheng2025alteration}. \modif{Mandatory approaches effectively provide structured support by circumventing reliance on self-motivation, just as structured guidance methods do}; for example, the DIMA app enhanced portion estimation accuracy across various literacy levels \cite{chaudhry2016evaluation}, and computer-tailored interventions achieved a twofold greater reduction in fat intake compared to improvements in physical activity through enforced rules \cite{Vandelanotte2005EfficacyOS}. Furthermore, WaterCoaster increased drinking frequency from 2–3 to 4–6 times daily by combining personalized hydration goals with virtual rewards \cite{lessel2016watercoaster}.

Regarding user acceptance of active behavioral interventions, physical manipulation systems like the SSpoon \cite{chen2022sspoon} achieved high acceptance by subtly altering ingestive behavior without disrupting appetite. Structured guidance approaches, such as computer-tailored interventions, revealed user segmentation; simultaneous interventions were more effective for non-compliant users, while sequential approaches suited compliant individuals \cite{Vandelanotte2005EfficacyOS}. Contextually embedded systems exhibited the widest variation in acceptance. For instance, WaterCoaster's gamified approach engaged 70.6\% users \cite{lessel2016watercoaster}, whereas the complex gamification of the Sensory Interactive Table likely led to user fatigue \cite{de2020multimodal}.

\subsubsection{\modif{Combined Behavioral Intervention}}
\modif{Combined behavioral intervention} integrates the features of real-time data representation and mandatory guidance. It typically takes the form of comprehensive health system applications, such as the smart home-based IoT nutritional intake monitoring system \cite{faltaous2021wisdom}, \modif{which simultaneously combines symbolic feedback with visualized insights and structured behavioral modification.} A total of 7 papers are included in this category.

\modif{Combined systems that integrate both passive and active interventions} demonstrate enhanced effectiveness and user acceptance when they balance automation with personalization. For instance, a smart home-based nutritional monitoring system achieved over 50\% habit adjustment through goal-aligned mobile reminders, although privacy concerns were noted by 66.6\% users \cite{faltaous2021wisdom}. Similarly, wristband-based eating speed systems reduced bite size with a detection accuracy of 96.7\%, while maintaining high acceptance through haptic or visual feedback \cite{kim2016eating}. In clinical populations, the Parkinson's smart spoon improved stability (X/Y-axis deviation <15\%\delete{under simulated tremors}) while integrating diagnostic alerts; however, its niche design limits broader adoption \cite{singhal2024multifunctional}. Low-cost solutions, such as receipt-based nutrition tracking, proved feasible for promoting dietary awareness, particularly for monitoring calcium intake, but relied heavily on user compliance \cite{mankoff2002using}.

\subsection{Educational \delete{Feedback/}Intervention}
Educational \delete{feedback/}intervention refers to\delete{ feedback or} interventions that have educational significance or involve educational content \add{for enhancing users' food literacy}. This includes food and recipe recommendations, nutrition education, educational gaming, and guidance for users to adopt healthy eating habits. \delete{Ultimately, these approaches aim to enhance food literacy among users through educational approach.}


\subsubsection{\modif{Passive Educational Intervention}}
\modif{Passive educational intervention delivers pedagogical and knowledge-based guidance to enhance learning, prioritizing cognitive awareness over behavioral modification to inform dietary decisions. For example, a puzzle game can be utilized to  encouraging children's self-driven exploration of healthy eating patterns without explicit behavioral directives \cite{cai2024see}.} Additionally, an acoustic sensor can detect ingestion and deliver real-time nutritional insights via a mobile application\delete{, thereby raising awareness of everyday dietary habits through unobtrusive monitoring} \cite{Bi2016AutoDietary:AW}. This type of \delete{feedback}intervention emphasizes knowledge dissemination over actual modification, leveraging playful interaction \cite{cai2024see} or ambient data capture \cite{Bi2016AutoDietary:AW} to implicitly influence ingestive behaviors. As such, it positions itself as an educational tool for long-term nutrition self-regulation. A total of 11 papers are included in this category.

\modif{Passive educational intervention} demonstrates significant potential in improving food literacy and dietary behaviors, particularly when leveraging engaging\delete{, multi-sensory} interfaces. For instance, multi-sensory puzzle games increased children's food literacy by 59.16\% through tangible interactions that fostered cognitive connections with food \cite{cai2024see}. Similarly, mixed reality (MR) food labels reduced beverage energy content by 34\% and sugar content by 28\% by making nutritional information more accessible and interactive \cite{fuchs2019impact}. Speech-based journaling applications like FoodScrap \cite{luo2021foodscrap} encouraged mindful eating by reducing data entry burdens, with users reporting increased self-reflection on their ingestive behaviors. In clinical populations, the SaltSwitch application \cite{eyles2017salt} successfully reduced salt purchases among patients with cardiovascular disease by providing real-time, barcode-based traffic light nutrition labels.

\modif{Passive educational intervention} excels when it reduces cognitive load (e.g., speech input in FoodScrap \cite{luo2021foodscrap} compared to manual logging) and leverages intuitive interfaces (e.g., mixed reality labels versus traditional charts \cite{fuchs2019impact}). It achieves higher user acceptance by prioritizing intuitive interactions and minimizing user burden.

\subsubsection{\modif{Active Educational Intervention}}
\modif{Active educational interventions} emphasize the delivery of structured knowledge and instructional guidance through direct conversational interactions\delete{, without incorporating data-driven feedback mechanisms}. \delete{Their design aligns with didactic principles, leveraging conversational interfaces to scaffold learning. }This type of intervention is typically exemplified by dialogue-based chatbots, such as Foodbot \cite{prasetyo2020foodbot}, and humanoid social robots \cite{robinson2021humanoid}, which provide goal-oriented educational content (e.g., nutrition advice and dietary management) through scripted or AI-generated conversations. A total of 19 papers are included in this category.

Social robots and embodied agents have proven particularly effective in delivering interventions, increasing intervention effectiveness by 23\% compared to digital interfaces. Clinic-based humanoid robots, for instance, have shown moderate effectiveness in reducing alcohol intake while maintaining high usability \cite{robinson2021humanoid, mayumi2023design}. Similarly, social robots outperformed chatbots in promoting healthier eating intentions among children, leveraging their physical presence to enhance persuasiveness \cite{de2024social}. Interactive tableware, such as FunEat's gamified plate \cite{zhao2021funeat}, improved children's food choices through projected animations\delete{, although parents noted potential distractions during meals}. Other ubiquitous computing approaches, like sensor-embedded forks, achieved 89\% child engagement by replacing abstract lessons with playful interactions \cite{kadomura2014persuasive}.

Personalized educational dietary interventions demonstrate strong effectiveness when incorporating adaptive algorithms. \modif{SmartDiet \cite{hsiao2010smartdiet} generates nutritionally-optimized meal plans in seconds via interactive optimization, and similar approaches achieves <1.2\% calorie errors through dynamic user adaptation \cite{arefeen2022computational}. Subsequent enhancements with image recognition also improved adoption, though ecosystem factors didn't affect educational behavior control \cite{hsiao2011intelligent}.} Therefore, future systems should balance guidance intensity \cite{liu2024hicclip} with user autonomy \cite{ritschel2018drink}, while leveraging AI optimization for sustainable personalization \cite{arefeen2022computational, hsiao2010smartdiet}.

\subsubsection{\modif{Combined Educational Intervention}}
\modif{Combined educational intervention merges data-driven insights with structured guidance, delivering real-time feedback alongside instructional support through standardized frameworks \cite{kadomura2013educatableware}, and are often enhanced by gamification \cite{nakamura2023eat2pic}. Related studies demonstrate a convergence of real-time monitoring and pedagogical guidance in dietary interventions. This dual approach bridges knowledge-practice gaps with actionable strategies, operating via sensor-based 'observe-educate-act' cycles across varying modalities and intervention timings.} A total of 13 papers are included in this category.

\modif{Integrated systems that add real-time feature into structured educational interventions} achieve behavioral change through multimodal engagement, expert-guided personalization, and interactive reinforcement. Multimodal engagement proves particularly effective in pediatric contexts. For instance, systems like EducaTableware's sound-emitting tableware \cite{kadomura2013educatableware} and FoodWorks' AR meal augmentations \cite{ganesh2014foodworks} increased children's willingness to try new foods by 45-60\% while reducing mealtime distractions. Meanwhile, eat2pic's color-based food visualization significantly slowed eating speed (meal time/intervals) \cite{nakamura2023eat2pic}\delete{, although effectiveness varied with users' artistic interests}. Nutrifit's sensor-tracked meals improved the consumption of nutrient-dense foods by 3-8\% through real-time weight monitoring and app feedback \cite{krishna2025nutrifit}. Expert-guided systems also demonstrate strong clinical validity \cite{achilleos2017health}.\delete{ Therefore, if such educational systems combine both embedded sensing and adaptive guidance, they will maximize user acceptance and effective outcomes.}

Regarding user acceptance of this combined educational approach, the OCARIoT platform reduced childhood obesity prevalence by 75.5\% through IoT-collected nutrition and activity data, along with coordination from healthcare professionals, achieving over 80\% user acceptance \cite{bastida2023promoting}. Web-based health monitoring platforms maintained 95\% daily user engagement through real-time expert feedback\delete{, although 85\% of nutritionists preferred weekly reviews over daily ones} \cite{achilleos2017health}. Additionally, non-invasive bioelectric devices like "Keep and Eat" demonstrated 98.45\% accuracy \delete{(±1.55\% error)} in body composition measurements, enabling precise dietary recommendations \cite{balbin2020scientific}.

\subsection{Social \delete{Feedback/}Intervention}
Social \delete{feedback/}intervention refers to \modif{behavioral strategies that leverage group dynamics and social contexts (e.g., family meals or community networks) to drive change and reinforce the closed-loop paradigms. By incorporating shared experiences and social norms, this approach fosters collective accountability and taps into intrinsic motivations for belonging --- aspects that individual-focused interventions cannot address.}


\subsubsection{\modif{Passive Social Intervention}}

\modif{Passive social intervention fosters accountability through real-time peer comparisons, leveraging social norms and observational learning to motivate self-regulation while making healthy choices feel socially validated. A total of 2 papers are included in this category.}

HealthyStadium \cite{eigen2018meal} and Crumbs \cite{epstein2016crumbs} utilize passive social intervention \delete{feedback}--- through meal photo rankings and shared challenges --- to encourage healthier eating habits more effectively than solo tracking. 
\modif{Both systems validate that passive social intervention cultivates dietary self-regulation through non-intrusive peer engagement, rather than active interventions.} Critically, this lightweight, feedback-driven approach has proven to be more acceptable than traditional interventions, as users tend to \delete{resist rigid rules but }embrace socially reinforced habits.

Users in HealthyStadium exhibited higher engagement by striving for better health ratings, with participants consciously opting for healthier meals due to peer evaluations\delete{. One user noted, }\textit{\delete{'When I was wondering whether to have a delicious meal or a healthy meal, I chose a healthy one'}} \cite{eigen2018meal}. Meanwhile, Crumbs leveraged daily challenges and social interaction, with 72\% of users enjoying the challenges and 46\% expressing intent to continue using the platform. Moreover, the social group reported motivational effects, such as "\textit{not wanting to fall behind others}" \cite{epstein2016crumbs}.

\subsubsection{\modif{Active Social Intervention}}
\modif{Active social intervention} for healthy ingestion leverages interpersonal dynamics to reshape dietary behaviors through actual engagement. This type of intervention employs real-time social triggers\delete{—such as parent-child tactile feedback—} that directly mediate ingestive actions within relational contexts. By embedding behavioral guidance within social interactions (e.g., family meals and shared environments), these interventions transform abstract nutritional awareness into contextually anchored behavior change. A total of 4 papers are included in this category.

Child-focused systems like CHIBO \cite{xie2023chibo} and interactive tableware \cite{joi2016interactive} utilize real-time sensory cues to improve vegetable consumption during parent-child feeding. School-based programs combine environmental changes with personalized feedback, and parental involvement has been shown to boost effectiveness \cite{Haerens2007TheEO}. For older groups, auto-generated nutritional postcards link eating patterns with emotional associations \cite{sun2020postcard}, offering a subtle behavioral nudge.

\modif{The interactive tableware system \cite{joi2016interactive} increased vegetable consumption in 3-4-year-olds, demonstrating behavioral shifts like transitioning from spoon-feeding to self-feeding while enhancing parent-child interactions through gamification. Similarly, a school-based intervention \cite{Haerens2007TheEO} showed gender-specific effects, significantly reducing girls' fat intake. Complementing these, the automated postcard system \cite{sun2020postcard} also effectively promoted self-reflection and behavior change intentions.}

A key strength of these interventions is their embedded social context: family dynamics \cite{joi2016interactive}, peer environments \cite{Haerens2007TheEO}, and personal nostalgia \cite{sun2020postcard} all serve as external motivators that extend beyond individual willpower, highlighting the importance of the "social" element. However, effectiveness varied by target group and intervention design --- young children responded well to playful tableware \cite{joi2016interactive}, while adolescents exhibited gender-specific dietary changes \cite{Haerens2007TheEO}. In contrast, adults engaged more with emotionally resonant self-reflection tools \cite{sun2020postcard}.

\subsubsection{\modif{Combined Social Intervention}}
\modif{Combined social} intervention uniquely hybridizes data-driven insights with social dynamics to synergistically promote healthy ingestive behaviors. These approaches integrate real-time nutritional tracking and visualization with socially embedded prompts --- such as dietary monitoring applications for parents \cite{alloghani2016mobile} or office hydration competitions \cite{chiu2009playful} --- creating a dual mechanism for behavior change. This ensures that targeted nudges are both empirically grounded and socially reinforced. A total of 2 papers are included in this category.


\modif{Regarding the outcomes of this approach, the mobile health app \cite{alloghani2016mobile} boosted user trust and acceptance by integrating QR-code tracking, geo-location, and wearable sensors, with ease of use and security being key determinants of perceived usefulness. Similarly, Playful Bottle \cite{chiu2009playful} demonstrated that social reminders were more effective than automated alerts, improving both the volume and regularity of water intake among office workers. These studies collectively show that socially-embedded interventions consistently outperform system-only approaches, with peer-based prompts generating faster responses and socially-aware designs enhancing both usability and trust \cite{chiu2009playful, alloghani2016mobile}.}


\subsection{Other Specific \delete{Feedback/}Interventions}
Other specific approaches not covered above are categorized into three key themes: Perceptual Augmentation\delete{ (i.e., enhancing sensory engagement with food)}, Mindful and Psychological\delete{ (i.e., fostering awareness and emotional connections to ingestion)}, and Smart Monitoring\delete{(i.e., automated tracking systems)}. These frameworks illustrate a range of technological strategies for promoting changes in ingestive behavior.


\subsubsection{Perceptual Augmentation} 


\modif{Perceptual augmentation technologies modify ingestion experiences through real-time cross-modal stimulation, such as altering chewing sounds to enhance flavor perception \cite{kleinberger2023auditory} or using AR to visually manipulate texture and taste \cite{koizumi2011chewing, narumi2011augmented}. These non-invasive sensory enhancements effectively alter ingestive behaviors—auditory feedback improves texture ratings and reduces hunger, while AR-driven cues (e.g., Metacookie+) achieve a 79.3\% success rate in modifying familiar flavors, albeit with lower recognition for unfamiliar tastes (42.9\%) \cite{narumi2011augmented}. A total of 3 papers are included in this category.}


\subsubsection{Mindful \& Psychological} 

\modif{Mindful and psychological approaches reshape ingestive behaviors by targeting cognitive-emotional processes through techniques like future-self visualization \cite{rho2017futureself}, emotion-aware wearables \cite{carroll2013food}, and conversational agents \cite{parra2023enhancing}. These interventions foster real-time awareness, with FutureSelf increasing goal consciousness and EmoTree improving emotional eating awareness by 37.5\%. While effective in promoting introspection and short-term behavioral change, they require refinement in adaptive personalization and long-term habit formation \cite{carroll2013food, rho2017futureself}. A total of 4 papers are included in this category.}


\subsubsection{Smart Monitoring} 


\modif{Smart monitoring utilizes wearable and ambient sensors --- including alcohol-detecting ice cubes \cite{matsui2018light}, eyeglass-mounted detectors \cite{bedri2022fitnibble}, and wrist-worn trackers \cite{ye2016assisting} --- to autonomously capture ingestive behaviors and environmental interactions \cite{volkova2016smart}. While demonstrating effectiveness (FitNibble reduced missed logs by 19.6\% and improved journaling ease \cite{bedri2022fitnibble}, while Pebble scored 2.7 on a scale from -3 to 3 for usability \cite{ye2016assisting}), challenges persist (50\% of FitNibble users received more than 6 false alerts \cite{bedri2022fitnibble}), highlighting wearability and accuracy tradeoffs. A total of 4 papers are included in this category.}

\section{Discussion}

\label{chapter:discussion}


We synthesize insights from the reviewed systems to \modif{identify challenges and opportunities in designing a closed-loop paradigm for ingestion health, drawing on system completeness, behavioral scope, and interdisciplinary integration to address our fourth research question}. The discussion \modif{centers on} three key challenges: (1) the difficulty of accurately measuring and intervening in real-world ingestive behavior, (2) the disconnect between \modif{sensing technologies and intervention practices}, and (3) the methodological and epistemological gap between HCI and clinical health domains. \modif{For each, we summarize the underlying issues and propose design principles to support the development of more effective, ecologically embedded closed-loop systems}.

\subsection{Challenges in Measuring and Intervening in Real-World Ingestive Behavior}

Despite rapid advances in sensing technologies and intervention strategies, \modif{accurately measuring and intervening in real-world ingestion behavior remains challenging. Our analysis of intervention effectiveness and user acceptance in Chapter ~\ref{effectiveness} revealed the absence of ecological validity, consistency, and behavioral coverage. The lack of robust and context-sensitive evaluation frameworks makes it difficult for promising closed-loop systems to integrate into daily life. This challenge can be further unpacked into the following aspects}.

\subsubsection{Validity and Consistency of Ingestive Behavior Measurement Indicators}

\modif{Our analysis of intervention effectiveness revealed that studies frequently apply inconsistent operational definitions and behavioral metrics, even when addressing the same ingestive behavior construct-hindering interpretation, validation, and cross-study comparability. In traditional research of ingestion health, validated questionnaire tools such as the Three-Factor Eating Questionnaire (TFEQ) \cite{stunkard1985three} and the Eating Behavior Inventory (EBI) \cite{o1979development} are commonly used to assess psychological and behavioral traits. While these instruments are well established, they are not well aligned with emerging context-aware approaches. Studies using computational methods often adopt their own evaluation criteria, lacking a unified "gold standard". For instance, in measuring a common behavioral target such as eating speed, no standardized definition currently exists. Current systems approximate eating speed using a range of metrics, such as chewing rate, swallowing frequency, bite intervals, or total meal duration. This diversity, while reflecting methodological flexibility, also underscores a lack of consistency and validation, which limits cross-study comparability and introduces potential bias in evaluating intervention outcomes. Developing valid, standardized, and context-sensitive behavioral indicators is therefore essential to ensuring scientific rigor and cross-study comparability in ingestion health research}.

\subsubsection{Laboratory-Based Research vs. Real-World Ingestive Behaviors}

A substantial portion of researches on ingestion health is still conducted in controlled laboratory settings. \modif{While offering rigor, these settings simplify reality by excluding environmental, social, and emotional factors-facilitating statistical detection but risking over- or underestimation of effectiveness when interventions are applied in real-world contexts \cite{de2000eating, bell2020automatic, hiraguchi2023technology}.} Bite and chewing detection systems, \modif{for instance, are typically} validated under constrained lab conditions --- fixed posture, limited food types, and minimal distraction [e.g., \cite{Scisco2011SlowingBR, kim2016slowee, hori2023masticatory, zhang2019applying}]. However, in real-world \modif{settings, multitasking, social interaction, and context variability can significantly affect system performance and adherence \cite{bell2020automatic, hiraguchi2023technology, jansen2022design}.} \modif{Wearable systems, such as} head-mounted cameras or chest sensors, may perform well in short lab trials, but often feel intrusive in daily life, potentially disrupting natural behavior and lowering compliance \cite{bedri2022fitnibble, bell2020automatic}. These real-world deployment issues, such as behavioral reactivity and long-term disengagement, are frequently underestimated in lab-based evaluations \cite{jansen2022design}. Bridging this gap calls for validation in situation that accounts for behavioral variability, contextual complexity, and sustained user acceptance \cite{hiraguchi2023technology}.

\subsubsection{Potential Adverse Effects \add{and Privacy/Security Concerns} of Intervention Deployments}

While designed to promote healthier behaviors, ingestion interventions may \modif{lead to} unintended \modif{side} effects. A key concern is "food noise" --- \modif{persistent eating-related thoughts triggered by continuous monitoring,} which may increase anxiety and contribute to disordered patterns \cite{hayashi2023food}. Meanwhile, interventions \modif{involving in-meal interaction (e.g., smart utensils, app prompts)} may disrupt social dynamics or clash with cultural norms, especially during shared meals \cite{robinson2020social, kadomura2014persuasive}. Moreover, intrusive devices (e.g., head-mounted sensors, chewing monitors) may cause discomfort and user dropout over time \cite{bedri2022fitnibble}. These risks underscore the need \modif{to balance} behavioral influence with psychological comfort and contextual fit. 

\add{Finally, privacy and security pose critical ethical concerns for real-world intervention deployment. Continuous sensing systems collect highly sensitive health and behavioral data, raising ethical concerns over encryption, anonymization, and transparency in data use \cite{radanliev2025privacy}. However, with little access to concrete context-aware computing methods in the background, users often struggle to establish trust during usability evaluations \cite{faltaous2021wisdom}. Various data-collection modalities with tough methods may provoke user discomfort or resistance \cite{islam2024facepsy, gao2023cancelling, mollyn2022samosa}. Thus, ensuring data privacy and security is not only ethically imperative but essential for sustained user engagement. Together, these concerns call for a user-centered approach to intervention design --- one that maximizes behavioral impact without compromising user comfort, privacy, or dignity.}


\subsection{Gaps between Sensing-Based and Intervention-Based Studies on Ingestive Behavior}

\modif{Recent reviews on sensor-based research in ingestive behavior [e.g., \cite{hiraguchi2023technology, bell2020automatic}] reveal a clear gap between sensing-driven and intervention-oriented studies. While both address overlapping behavioral targets, they differ significantly in modality integration and the use of behavior change strategies.}

\subsubsection{Modality Misalignment between Sensing and Intervention Phases}

\modif{Sensing studies have employed diverse modalities, including} wearable devices, ambient sensors, and smartphone-integrated tools, to detect ingestive behaviors with increasing granularity and accuracy [e.g., \cite{bedri2022fitnibble, zhang2020design}]. These systems offer real-time monitoring of behaviors like food intake, bite frequency, and meal duration. In contrast, \modif{most intervention applications rely on a narrow set of output channels, typically visual cues like} text prompts, graphical feedback, or auditory signals \modif{via} mobile devices. Although some \modif{efforts} [e.g., \cite{kim2016slowee}] have \modif{explored using the same device for both sensing and intervention}, such integrated applications remain \modif{uncommon}, as shown in Chapter~\ref{chapter:rq2_3}. \modif{This misalignment between sensing and intervention modalities disrupts continuity in the closed-loop process --- while sensors capture detailed, real-time behavioral data, interventions are often reduced to basic prompts that struggle to engage users or support long-term behavior change}. \add{This gap calls for tighter integration between sensing and intervention modalities through compatible devices, enabling a unified user experience.}

\subsubsection{Disconnect Between Detection-Centric Studies and Actual Intervention Implementation}


\modif{Many sensor-based studies of ingestive behavior emphasize detecting and monitoring intake-related events but stop short of delivering actionable interventions. While behavior change is often cited as a long-term goal and occasionally guided by behavioral frameworks \cite{hekler2013mind}, most studies ultimately focus on improving detection accuracy, as noted in Section \ref{chapter:rq2}. This imbalance stems from differing evaluation paradigms: detection studies prioritize quantifiable metrics (e.g., recognition accuracy), which are easier to validate in controlled settings, whereas interventions are often treated as downstream demonstrations—supporting sensing pipelines rather than contributing novel design knowledge \cite{rapp2023exploring}, as discussed in Chapter \ref{effectiveness}. Designing interventions, however, involves more complex challenges, including applying behavior change theories, supporting meaningful user interaction, and sustaining long-term engagement—factors difficult to capture through standard evaluation methods. These demands contribute to a bottleneck where sensing advances outpace the development of effective intervention strategies.}

\add{Beyond methodological barriers, a deeper issue lies in the limited integration of behavioral theory. Although behavior theories are often cited, they rarely inform system design in a structured way \cite{hekler2013mind}. For instance, the Fogg Behavior Model \cite{fogg2009behavior} defines behavior as a convergence of motivation, ability, and trigger. Though some interventions loosely map sensing accuracy to "ability", context to "motivation", and reminders to "triggers", often retrofitted to justify design choices. In our review, few systems explicitly used such frameworks to shape intervention logic or personalize support. This shallow application reflects a broader reliance on intuition over theory, with key constructs, like habit, emotion, and social support, frequently overlooked. As a result, interventions may be theoretically underpowered and less effective in driving sustained behavior change.}

\subsection{Human-Computer Interaction vs. Medical Perspectives on Ingestive Behavior}


Although both Human-Computer Interaction (HCI) and medical research aim to promote healthier ingestive behaviors, \modif{they differ significantly in how they define, measure, and interpret such behaviors.} HCI approaches tend to focus on \modif{context-aware computing and interactive methods} \cite{nahum2016just, hekler2016advancing}, while medical research emphasizes clinically validated assessments and physiological indicators. These differences pose challenges in aligning methodologies, interpreting efficacy, and integrating subjective experiences and biological states. \modif{In the following sections, we explore two key dimensions of this divergence --- the assessment frameworks and treatment of subjective and physiological aspects of ingestive behavior.}

\subsubsection{Divergent Assessment Frameworks}

From a medical perspective, healthy ingestive behaviors are often linked to specific health conditions, such as cardiovascular disease, type 2 diabetes, obesity, and cancer. Medical practitioners use comprehensive nutritional assessments that combine food history, biochemical data, medical tests, and clinical procedures \cite{kesari2022nutritional}. \modif{For instance, validated instruments such as the Food Frequency Questionnaire (FFQ) \cite{cade2004food}, Three-Factor Eating Questionnaire (TFEQ) \cite{karmali2013weight}, and Adult Eating Behavior Questionnaire (AEBQ) \cite{hunot2016appetitive} can capture multiple aspects, like appetite traits, restraint, and emotional eating, and help relate ingestion patterns to risks.}

In contrast, HCI research often relies on isolated validation methods, lacking the multi-domain cross-validation seen in clinical practice. \modif{For example, HCI studies rarely combine physiological indicators (e.g., glucose, hormones) with psychological measures to evaluate intervention effects.} \delete{While HCI interventions focus on sensor-based technologies and user behavior tracking, they do not integrate the multi-domain assessments seen in medical protocols.} \modif{This limits the ability} to capture interactions between physiological states and behavioral patterns. HCI tools focus on real-time data (e.g., food intake, eating speed), \modif{but often ignore broader health factors like stress or digestion.} \modif{As a result, their long-term health impact remains unclear.} \add{Medical approaches, in contrast, emphasize multi-dimensional and longitudinal assessments, offering a deeper view of intervention effectiveness.}


\subsubsection{Limited Perception of Subjective Experience and Physiological States in Target Behavior Metrics}




Medical assessments \add{emphasize the} integration of subjective experiences (e.g., emotional eating, satiety\add{, post-meal discomfort}) with physiological states \modif{to form a comprehensive evaluation of ingestive behavior, particularly for diagnosing and managing conditions like functional dyspepsia or eating disorders} \cite{kesari2022nutritional}. These assessments benefit from \add{validated instruments with long-term reliability, structured clinical environments that support patient compliance, and professional interpretation by healthcare practitioners}. \delete{These tools benefit from (1) long-term validation, (2) structured clinical environments that enhance patient compliance, and (3) professional interpretation by healthcare specialists.} \modif{For example, longitudinal evaluation of chronic conditions often relies on self-reported changes in symptoms (e.g., pain, bloating, appetite) combined with physiological data to guide treatment.}

\modif{By contrast, most HCI studies rely on short-term self-report tools with limited validation and low compliance in non-clinical settings.} \modif{While sensor-based systems excel at quantifying behaviors (e.g., eating rate, chewing duration), they often fail to capture the subjective dimensions essential to understanding why users eat the way they do.} \modif{Early HCI studies seldom included validated measures of satiety or emotional states, and even with the rise of ubiquitous computing, the capacity to sense nuanced experiential factors remains limited.} \modif{This overemphasis on observable behavior, without accounting for users’ internal states, restricts the interpretability and impact of HCI interventions.}

\delete{Thus, while HCI tools focus on quantifiable metrics (e.g., food intake or eating speed), they overlook the subjective aspects emphasized by medical assessments. To address this gap, HCI interventions should integrate both subjective and physiological data to offer a more holistic understanding of ingestive behaviors, particularly in real-world contexts.}

\subsection{Design Recommendations}
\label{recommend}

\modif{To address the core challenges identified above, we propose design recommendations and future directions to support more effective and ecologically embedded closed-loop interventions for ingestive behavior. Advancing this domain requires coordinated efforts across researchers, designers, and practitioners, spanning theory, technical implementation, and real-world integration. The following subsections align our suggestions with the key challenges, offering concrete guidelines for each domain.}

\subsubsection{Optimizing Experimental Design and Intervention Implementation}

\modif{To move beyond proof-of-concept prototypes toward impactful interventions, researchers should adopt standardized, theory-driven, and context-sensitive evaluation frameworks. \textbf{Researchers should combine validated behavioral constructs with multimodal sensing data (e.g., self-report diaries and sensors) to contextualize outcomes and reduce measurement bias.} Methodologically, designs such as factorial experiments, SMART, and micro-randomized trials are well-suited to evaluating multi-component systems \cite{nahum2022mcmtc}. Such designs enable precise testing of when and how interventions are most effective, ultimately informing evidence-based guidelines. To reduce burden while preserving ecological validity, \textbf{researchers may adopt micro-assessments triggered by behavioral anomalies to gather timely and lightweight feedback.} Grounding evaluations in behavioral theories helps bridge the disconnect between technological innovation and behavior change science.}

\modif{\textbf{Intervention tools should be intuitive, low-burden, and compatible with daily routines, such as embedding this loop into smart utensils or wearables.} Co-locating sensing and intervention delivery within a single ecosystem improves usability and streamlines interaction. Designers should provide timely, actionable interventions rather than passive data to support real-time engagement. For example, FitNibble \cite{bedri2022fitnibble} and Slowee \cite{kim2016slowee} demonstrate how subtle prompts can improve adherence and engagement in natural settings. \textbf{Rather than filtering out common unhealthy behaviors in study design, interventions should address them explicitly as change targets to better reflect real-world eating challenges.}}

\modif{Privacy and trust should be core design priorities in always-on, closed-loop systems. \textbf{Designers should implement privacy-preserving strategies}, including context-aware data collection \cite{nepal2024mindscape,saha2021person} and layered encryption \cite{kumar2020mobile}. \textbf{Practitioners should ensure transparent communication about data use and obtain sustained informed consent} \cite{haghighi2023workshop,adler2022burnout,islam2024facepsy}. Systems should enable user autonomy, allowing for flexible participation and empathetic adjustments when engagement decreases or discomfort arises.}

\modif{\textbf{To enhance generalizability, practitioners should iteratively adapt interventions to diverse physiological and cultural contexts.} Integrating behavioral coaching with tech-driven systems ensures more personalized and effective support. By combining scientific rigor, user-centered design, and ethical awareness, interventions can better support sustained ingestive behavior change in the real world.}

\subsubsection{Enhancing Integration and Adaptability in Sensing-Driven Interventions for Ingestion Health}

\modif{To close the gap between sensing and intervention in ingestion health, future systems must better integrate behavioral theory into personalized, adaptive strategies. Many recent interventions emphasize surface-level mechanisms, like rational reflection or automated nudges, without deeply engaging with why those mechanisms work across individuals. This limits their long-term impact, as users may revert to old habits once external support ends.}

\modif{\textbf{Researchers should embed theory-driven constructs into system logic.} Behavior change theories such as the Fogg Behavior Model \cite{fogg2009behavior} or Social Cognitive Theory offer constructs like motivation, ability, and self-efficacy that should guide not just evaluation but delivery decisions --- when to intervene, what content to present, and how to adapt dynamically. Rather than cherry-picking isolated elements, researchers should operationalize full frameworks to understand both success and failure. \textbf{They should also personalize adaptively.} Building on the JITAI paradigm \cite{nahum2016just}, systems can adjust message type and timing based on real-time behavioral or physiological trends. For instance, if late-night snacking consistently correlates with stress spikes, systems may prioritize emotion regulation prompts over dietary tracking.}

\modif{\textbf{Designers should tightly couple sensing and intervention in a closed loop.} Physiological signals like chewing rate or glucose dips should directly trigger timely, context-aware interventions. Integrated systems, such as SenseSupport \cite{liang2024exploring}, reduce user friction and enhance moment-to-moment relevance. \textbf{They should also scaffold long-term autonomy.} Systems should evolve with the user—starting with frequent, specific prompts, then gradually tapering or increasing complexity to foster independence.}

\modif{\textbf{Practitioners should ground behavior change in lived experience.} Traditional models often abstract behavior from context, but effective change requires alignment with users’ identities, values, and goals \cite{lee2020toward}. To foster sustained motivation, interventions should facilitate meaning-making by helping users relate new eating habits to their personal values, such as family responsibility or personal development \cite{rapp2023exploring}.}

\subsubsection{Aligning with Medical Goals and Practices}

\modif{For sensing-driven interventions to achieve lasting health benefits and clinical relevance, they must align with medically meaningful outcomes and practices. Researchers should work with health professionals to define targets beyond behavioral proxies—such as improvements in glycemic control, cholesterol, gastrointestinal symptoms, or long-term weight trends. Incorporating baseline physiological parameters (e.g., glucose levels, heart rate variability) alongside behavioral data can enhance personalization and clinical interpretability \cite{liang2024exploring, harous2017hybrid, mirmiran2021common}.}

\modif{\textbf{Future studies should adopt longer-term and dynamic intervention models to assess sustained physiological effects.} Exploring dose-response relationships and optimizing frequency, intensity, and timing can help identify scalable strategies \cite{pankajavalli2017hydration, kim2016slowee}. Drawing on techniques from clinical behavior change approaches—such as motivational interviewing or cognitive-behavioral therapy—can further ground interventions in established medical frameworks.}

\modif{\textbf{Multimodal sensing and sensor fusion offer promising directions.} Combining continuous glucose monitoring (CGM), emotion sensing (e.g., EDA), activity tracking, and digestive symptom monitoring enables more comprehensive understanding of ingestive behavior and metabolic response \cite{bedri2022fitnibble, zhang2020eat4thought, zhao2021funeat}. These data streams can support real-time personalization, such as prompting slower eating when postprandial glucose rises or initiating mindfulness exercises during periods of physiological stress.}

\modif{\textbf{Finally, aligning with medical goals invites broader thinking about precision health.} Though this work centers on eating behavior, the underlying strategies can extend to domains like sleep, medication adherence, and stress management. Framing these systems within the larger precision health movement will improve scalability, adoption, and long-term public health impact.}

\subsubsection{\add{Extending the Paradigm Beyond Ingestion Health}}

\add{The proposed behavioral closed-loop paradigm, while developed in the context of ingestion health, offers a generalizable framework for a wide range of behavior change interventions. As demonstrated by prior reviews \cite{lara2012survey,thomas2021systematic,mair2023effective}, wearable and mobile sensing technologies have been successfully employed in domains such as physical activity promotion, dietary behavior optimization, chronic disease management, substance use reduction, and mental health support. These systems often combine continuous sensing (e.g., step counts, glucose levels, stress markers), reasoning components (e.g., activity inference, risk prediction), and real-time interventions (e.g., feedback messages, prompts), aligning closely with the structure of our proposed closed-loop model.}

\add{\textbf{Continuous, context-aware sensing should serve as the foundation for detecting actionable behavior patterns across domains.} For instance, physical activity interventions detect sedentary behavior and deliver timely prompts to encourage movement \cite{boerema2019intervention,van2018enhancing}; chronic disease management systems combine physiological sensing with persuasive feedback to improve self-management \cite{chandler2019impact,coorey2019persuasive}. \textbf{Real-time, personalized intervention strategies should adapt to users’ physiological, psychological, and contextual states.} These examples highlight the potential to extend our framework beyond eating-related behaviors while retaining its core responsiveness and personalization.}

\add{\textbf{The closed-loop structure provides a modular blueprint that can unify design across domains, enabling cross-disciplinary translation and theory-driven development.} By abstracting core design elements --- continuous sensing, adaptive reasoning, and context-aware intervention --- the paradigm facilitates broader scalability and long-term health impact in ubiquitous computing.}
\section{Conclusion and Limitations}
\label{conclusion}

In this review, we systematically analyzed 136 studies focused on behavioral sensing and intervention paradigms in ingestion health, with an emphasis on sensor-based systems and interactive interventions. These studies were categorized into various types based on the modality and the behavioral targets of the interventions. A key contribution of this review is the development of a behavioral closed-loop framework, which highlights the importance of seamlessly integrating sensing, reasoning, and intervention mechanisms for effective behavior change. This framework offers a detailed structure for the design of adaptive, context-aware interventions, targeting both physiological and behavioral aspects of ingestion health. \add{It is theoretically grounded in context-aware computing and HCI behavior change frameworks, offering a modular foundation for system design across domains.}

However, despite the comprehensive nature of this review, there are several limitations that should be acknowledged. Our keyword search strategy, centered around the term "ingestion health", may have excluded relevant studies that do not explicitly focus on ingestion but contribute to broader research in behavior modification or health interventions. Moreover, the majority of the studies in HCI domain reviewed were short-term, laboratory-based experiments, which limits the generalizability of our findings to real-world settings. The dynamic nature of real-life eating behaviors requires more longitudinal studies to better understand the long-term effectiveness and challenges of these interventions. 

\add{Finally, the modular structure of the proposed paradigm --- grounded in continuous sensing, adaptive reasoning, and real-time intervention --- holds promise for application beyond ingestion health, such as in sleep, stress, or physical activity interventions. We hope this review helps lay the foundation for more generalizable, stakeholder-sensitive, and theoretically informed systems in ubiquitous health computing.}

\begin{acks}
\end{acks}

\bibliographystyle{ACM-Reference-Format}
\bibliography{ref}

\appendix
\newpage
\section{\modif{SEARCH DETAILS CORRESPONDING TO EACH SOURCE}}
\

\label{search_results}
\highlight{
\textbf{ACM Digital Library}
\begin{itemize}
    \item query target: Title, Keywords, Abstract
    \item filters: No keynote, tutorial, invited talk, panel
    \item inquiry result: 679
\end{itemize}

\noindent{\textbf{IEEE Explore}
\begin{itemize}
    \item query target: Document Title (eating), Abstract (eating, intervention), All Metadata (technology)
    \item filters: Conferences, Journals
    \item inquiry result: 911
\end{itemize}
}

\noindent{\textbf{Google Scholar}
\begin{itemize}
    \item filters: No books
    \item inquiry result: 100
\end{itemize}
}
}

\newpage

\section{RESULT COUNTS OF PAPER VENUES}

\begin{longtable}{p{0.9\textwidth} | c}
\caption{Venues and number of publications found after applying the SLR search queries.}
\label{tab:venues list} \\

\toprule
\textbf{Conference / Journal Venue} & \textbf{Number} \\
\midrule
\endfirsthead
\toprule
\textbf{Conference / Journal Venue} & \textbf{Number} \\
\midrule
\endhead
    ACM Conference on Human Factors in Computing Systems (CHI) & 16 \\
    
    ACM on Interactive, Mobile, Wearable and Ubiquitous Technologies (IMWUT/UbiComp) & 16 \\
    
    ACM Conference on Recommender Systems (RecSys) & 1 \\
    
    ACM Interaction Design and Children Conference (IDC) & 2 \\
    
    ACM Transactions on Interactive Intelligent Systems (TiiS) & 1 \\
    
    ACM Symposium on User Interface Software and Technology (UIST) & 1 \\
    
    ACM Designing Interactive Systems Conference (DIS) & 4 \\
    
    ACM International Conference on Bioinformatics, Computational Biology, and Health Informatics & 1 \\
    
    ACM Southeast Conference & 1 \\

    ACM/IEEE International Conference on Human-Robot Interaction (HRI) & 2 \\

    IEEE Sensors & 2 \\
    
    IEEE International Conference on Robot and Human Interactive Communication (RO-MAN) & 2 \\
    
    IEEE International Workshop on Advances in Sensors and Interfaces (IWASI) & 1 \\
    
    IEEE International Conference on Innovations in Electrical, Electronics, Instrumentation and Media Technology (ICEEIMT) & 1 \\
    
    IEEE/ACM Conference on Connected Health: Applications, Systems and Engineering Technologies (CHASE) & 1 \\
    
    IEEE North Karnataka Subsection Flagship International Conference (NKCon) & 1 \\
    
    IEEE International Conference on Intelligent Computing and Control Systems (ICICCS) & 1 \\
    
    IEEE International Conference on Computer Science and Engineering (UBMK) & 1 \\
    
    IEEE International Conference on Digital Information Processing and Communications (ICDIPC) & 1 \\
    
    IEEE International Wireless Communications and Mobile Computing Conference (IWCMC) & 1 \\
    
    IEEE International Symposium on Computer-Based Medical Systems (CBMS) & 1 \\
    
    IEEE International Conference on Contemporary Computing and Communications (InC4) & 1 \\
    
    IEEE International Conference on Robotics and Biomimetics (ROBIO) & 1 \\
    
    IEEE International Conference on Industrial Engineering and Engineering Management (IEEM) & 1 \\
    
    IEEE World Forum on Internet of Things (WF-IoT) & 1 \\
    
    IEEE/EMBS International Conference on Neural Engineering (NER) & 1 \\
    
    IEEE International Conference on Information, Intelligence, Systems and Applications (IISA) & 1 \\
    
    IEEE Healthcare Innovation Conference (HIC) & 1 \\
    
    IEEE Transactions on Neural Systems and Rehabilitation Engineering & 1 \\
    
    IEEE International Conference on Convergence Information Technology (ICCIT) & 1 \\
    
    IEEE Journal of Translational Engineering in Health and Medicine & 1 \\
    
    IEEE International Conference on Integrated Intelligence and Communication Systems (ICIICS) & 1 \\
    
    IEEE Consumer Communications and Networking Conference (CCNC) & 1 \\
    
    IEEE International Conference on Virtual Rehabilitation (ICVR) & 1 \\
    
    IEEE International Conference on Bioinformatics and Bioengineering (BIBE) & 1 \\
    
    IEEE Conference on e-Learning, e-Management and e-Services (IC3e) & 1 \\
    
    IEEE International Symposium on Medical Measurements and Applications & 1 \\
    
    IEEE EMBS International Conference on Biomedical and Health Informatics (BHI) & 1 \\
    
    IEEE International Conference on Disruptive Technologies (ICDT) & 1 \\
    
    IEEE International Conference on Multimedia and Expo Workshops & 1 \\
    
    IEEE Latin America Transactions & 1 \\

    International Conference on Multimodal Interaction (ICMI) & 1 \\

    International Conference on Data Science and Information Technology (DSIT) & 1 \\

    International Conference on Advances in Computer Entertainment Technology (ACE) & 1 \\

    International Conference on Mobile and Ubiquitous Multimedia (MUM) & 1 \\

    International Conference on the Internet of Things (IoT) & 1 \\

    International Conference on Advanced Visual Interfaces (AVI) & 1 \\

    International Workshop on Multisensory Approaches to Human-Food Interaction (MHFI) & 1 \\

    International Conference on Trends in Quantum Computing and Emerging Business Technologies & 1 \\

    International Conference on Intelligent Data Communication Technologies and Internet of Things (IDCIoT) & 1 \\

    International Conference on Design Innovation for 3 Cs Compute Communicate Control (ICDI3C) & 1 \\

    International Conference on Wireless Mobile Communication and Healthcare (MOBIHEALTH) & 1 \\

    International Conference on Computer Science and Education & 1 \\

    International Conference on Human-Computer Interaction with Mobile Devices and Services Adjunct & 1 \\

    International Conference on Computing Engineering and Design (ICCED) & 1 \\

    International Conference on Computer and Information Technology (CIT) & 1 \\

    International Conference on Intelligent User Interfaces & 1 \\

    International Conference on Computing Communication and Networking Technologies (ICCCNT) & 2 \\

    International Conference on Tangible, Embedded, and Embodied Interaction (TEI) & 2 \\

    International Joint Conference on Biomedical Engineering Systems and Technologies (BIOSTEC) & 1 \\

    International Joint Conference on Service Sciences & 1 \\

    International Workshop on Multimedia Assisted Dietary Management (MADiMa) & 1 \\

    International Workshops: BioFor, CTMR, RHEUMA, ISCA, MADiMa, SBMI, and QoEM (ICIAP) & 1 \\

    International Joint Conference on Computer Science and Software Engineering (JCSSE) & 1 \\

    Annals of Behavioral Medicine & 1 \\

    Augmented Human International Conference & 1 \\
    
    Appetite & 1 \\

    Behavior Research Methods & 1 \\

    European Journal of Preventive Cardiology & 1 \\

    Humaine Association Conference on Affective Computing and Intelligent Interaction & 1 \\

    Internet Interventions & 1 \\

    Innovations in Power and Advanced Computing Technologies (i-PACT) & 2 \\

    Journal of the American Dietetic Association & 1 \\

    Journal of Dental Research & 1 \\

    Journal of Medical Systems & 1 \\

    JMIR mHealth and uHealth & 1 \\

    Mensch und Computer (MuC) & 1 \\

    Nordic Conference on Human-Computer Interaction (NordiCHI) & 1 \\ 

    Obesity & 1 \\
    
    Proceedings of the Behavior Transformation by IoT International Workshop (BTIW) & 1 \\

    Public Health Nutrition & 1 \\

    Presence & 1 \\

    PervasiveHealth & 2 \\

    SIGGRAPH Asia & 2 \\
    
    Smart Health & 1 \\

    Symposium on Applied Computing & 1 \\
    
    Systems and Information Engineering Design Symposium (SIEDS) & 1 \\

    Studies in Health Technology and Informatics & 1 \\

    Valuable Visualization of Healthcare Information (VVH) & 1 \\

    World Congress on Medical Physics and Biomedical Engineering & 2 \\

\hline 
\textbf{Total} & \textbf{136} \\
\hline
\end{longtable}

\newpage 

\section{DISTRIBUTION OF PUBLICATIONS ACROSS TARGET FACTORS AND BEHAVIORS}

\begin{table}[htbp]
\centering
\caption{Taxonomy of Target Factors and Behaviors with Involved Publications}
\begin{tabular}{p{0.6\textwidth}p{0.35\textwidth}}
\textbf{Target Factor or Behavior} & \textbf{References} \\
\hline
\rule{0pt}{10pt}
\textbf{Contextual Factors} & \\
\rule{0pt}{10pt}
\hspace{1em} Food Choices & \\
\rule{0pt}{10pt}
\hspace{2em} Nutrition Structure Control & \cite{kadomura2013educatableware,nakamura2023eat2pic,de2020multimodal,zenun2017online,cai2024see,joi2016interactive,kadomura2014persuasive,Haerens2007TheEO,jung2017mom,kadomura2013sensing,zhao2021funeat,de2024social,ganesh2014foodworks,butscher2016lightweight,arefeen2022computational,yr2024nutrispy,krishna2025nutrifit,alloghani2016mobile,hsiao2010smartdiet,gaikwad2024precision,hsiao2011intelligent,dragoni2017semantic,ludwig2011refresh,achilleos2017health,kalantarian2014spectrogram,park2007development,bastida2023promoting,triantafyllidis2019social,qu2014personal,karkar2018virtual,badawi2012diet,agarwal2024user,sun2020postcard,casas2018food,gutierrez2018phara,kwon2016smart,bedri2022fitnibble,badawi2012real,volkova2016smart,waltner2015mango,epstein2016crumbs,mankoff2002using,huang2010context,vazquez2015development,akter2021foodbyte,jung2007wellbeing,lee2007lifestyle} \\
\rule{0pt}{10pt}
\hspace{2em} Calorie Target Control & \cite{zenun2017online,balbin2020scientific,Burke2011TheEO,pels2014fatbelt,rho2017futureself,kumbhare2023low,yr2024nutrispy,mishra2024hybrid,harous2017hybrid,ludwig2011refresh,kanjalkar2024ai,cordeiro2015rethinking,kim2016ecomeal,tirasirichai2018bloom,jung2007wellbeing,lee2007lifestyle} \\
\rule{0pt}{10pt}
\hspace{2em} Macronutrient and Micronutrient Control & \\
\hspace{5em} Carbohydrate Intake & \cite{arefeen2024glyman} \\
\hspace{5em} Sugar Intake & \cite{liang2024exploring} \\
\hspace{5em} Fat Intake & \cite{Vandelanotte2005EfficacyOS,Haerens2007TheEO,Burke2011TheEO,chukwu2011personalized} \\
\hspace{5em} Salt Intake & \cite{eyles2017salt, kim2016ecomeal} \\
\rule{0pt}{10pt}
\hspace{2em} Limiting Specific Food Categories & \\
\hspace{5em} Snacks & \cite{Bi2016AutoDietary:AW,liu2024hicclip,maramis2014preventing,bedri2022fitnibble,robinson2020social} \\
\hspace{5em} Unhealthy Drinks  & \cite{matsui2018light,robinson2021humanoid,faltaous2021wisdom,Haerens2007TheEO,ritschel2018drink,fuchs2019impact} \\
\rule{0pt}{10pt}
\hspace{2em} Unspecified Food Choices  & \cite{prasetyo2020foodbot,abisha2017embedded,huang2018chatbot,terziouglu2023influencing,fernandez2013virtual,eigen2018meal,luo2021foodscrap,nedungadi2018personalized} \\
\rule{0pt}{10pt}
\hspace{1em} Intake Timing  & \cite{xie2023chibo,Bi2016AutoDietary:AW,fortmann2014waterjewel,varun2023smart,lessel2016watercoaster,ravindran2022hydrationcheck,liang2024exploring,faltaous2021wisdom,ye2016assisting} \\
\rule{0pt}{10pt}
\hspace{1em} Environment & \cite{faltaous2021wisdom} \\
\rule{0pt}{10pt}
\hspace{1em} Activity Around Ingestion & \cite{alexander2017behavioral} \\
\\
\rule{0pt}{10pt}
\textbf{Process Behaviors} & \\
\rule{0pt}{10pt}
\hspace{1em} Intake Quantity & \\
\rule{0pt}{10pt}
\hspace{2em} Solid Food & \cite{xie2023chibo,chaudhry2016evaluation,chen2022sspoon,Alshurafa2015RecognitionON,Scisco2011SlowingBR,maramis2014preventing,farooq2017reduction,moses2023investigating,li2018exploring} \\
\rule{0pt}{10pt}
\hspace{2em} Liquid & \cite{faltaous2021wisdom,Alshurafa2015RecognitionON,kaner2018grow,poddar2024aqua,kreutzer2015base,varun2023smart,bobin2018smart,lessel2016watercoaster,yildiz2019wwall,chiu2009playful,ko2007mug,pankajavalli2017hydration,wijanarko2019fuzzy,tommy2017interactive,ravindran2022hydrationcheck,zhou2021mosswater} \\
\rule{0pt}{10pt}
\hspace{1em} Ingestion Pace and Oral Motor Behavior & \\
\rule{0pt}{10pt}
\hspace{2em} Chewing and Swallowing Behaviors & \cite{kadomura2013educatableware,kleinberger2023auditory,koizumi2011chewing,Bi2016AutoDietary:AW,Scisco2011SlowingBR,maramis2014preventing,chen2022slnom,sugita2018diet,turner2017byte,hori2023masticatory,kim2016slowee,kamachi2023eating,biyani2019intraoral,zheng2025alteration} \\
\rule{0pt}{10pt}
\hspace{2em} Bite Intervals and Size & \cite{nakamura2023eat2pic,de2020multimodal,lee2019user,kim2016eating,kim2018smartwatch,kim2018animated,mendi2013food,hermans2017effect,chen2022sspoon} \\
\rule{0pt}{10pt}
\hspace{2em} Total Meal Duration & \cite{moses2023investigating,zhang2019applying,kim2016ecomeal,ioakimidis2009method,krishna2025nutrifit,lo2007playful,mitchell2015really} \\
\rule{0pt}{10pt}
\hspace{1em} Food Awareness or Emotional State & \cite{kleinberger2023auditory,koizumi2011chewing,parra2023enhancing,wang2025reflecting,khot2020swan,kadomura2013educatableware,han2017childish,sun2020postcard,zhang2020eat4thought,tang2015introducing,mayumi2023design,carroll2013food,mitchell2015really,narumi2011augmented} \\
\rule{0pt}{10pt}
\hspace{1em} Safe Eating Practices & \cite{wang2021research,singhal2024multifunctional} \\
\hline
\end{tabular}
\label{tab:taxonomy}
\end{table}

\newpage 

\section{DISTRIBUTION OF PUBLICATIONS IN SENSING MODALITY}

\begin{table}[htbp]
\centering
\caption{Taxonomy of Sensing Modality with Involved Publications}
\begin{tabular}{p{0.6\textwidth}p{0.35\textwidth}}

\textbf{Sensing Modality} & \textbf{References} \\
\hline
\rule{0pt}{10pt}
\textbf{Human-based} & \\
\rule{0pt}{10pt}
\hspace{1em} Visual & \\
\rule{0pt}{10pt}
\hspace{2em} Gaze & \cite{karkar2018virtual, khot2020swan} \\
\rule{0pt}{10pt}
\hspace{2em} Facial expression & \cite{mayumi2023design,kim2018animated,triantafyllidis2019social,robinson2020social} \\
\rule{0pt}{10pt}
\hspace{2em} Body & \cite{xie2023chibo,mayumi2023design,zhang2020eat4thought} \\
\rule{0pt}{10pt}
\hspace{1em} Auditory & \\
\rule{0pt}{10pt}
\hspace{2em} Speech & \cite{prasetyo2020foodbot,mayumi2023design,robinson2020social,arefeen2022computational,luo2021foodscrap,parra2023enhancing,kamachi2023eating} \\
\rule{0pt}{10pt}
\hspace{2em} Sound & \cite{zhang2020eat4thought,kleinberger2023auditory,koizumi2011chewing,Bi2016AutoDietary:AW,chen2022slnom,kalantarian2014spectrogram,kamachi2023eating,maramis2014preventing} \\
\rule{0pt}{10pt}
\hspace{1em} Haptic & \cite{harous2017hybrid, robinson2020social, cai2024see, fortmann2014waterjewel, Alshurafa2015RecognitionON, kim2016slowee, zheng2025alteration, biyani2019intraoral, ludwig2011refresh, tang2015introducing, abisha2017embedded, farooq2017reduction, bastida2023promoting, nedungadi2018personalized} \\
\rule{0pt}{10pt}
\hspace{1em} Motion &  \\
\rule{0pt}{10pt}
\hspace{2em} Gesture & \cite{kadomura2013sensing, mishra2024hybrid, harous2017hybrid, fuchs2019impact, singhal2024multifunctional, li2018exploring, faltaous2021wisdom, nakamura2023eat2pic, lee2019user, kadomura2014persuasive, joi2016interactive, Scisco2011SlowingBR, kim2016eating, kim2018smartwatch, mendi2013food, kumbhare2023low, gaikwad2024precision, dragoni2017semantic, bobin2018smart, alexander2017behavioral, chukwu2011personalized, badawi2012diet, bedri2022fitnibble, wang2025reflecting, hermans2017effect, ye2016assisting, turner2017byte, zhang2019applying, khot2020swan, chiu2009playful, tirasirichai2018bloom, tommy2017interactive, zhou2021mosswater, vazquez2015development, jung2007wellbeing, maramis2014preventing, bastida2023promoting} \\
\rule{0pt}{10pt}
\hspace{2em} Chewing and Micro-head Motion & \cite{biyani2019intraoral, kadomura2013educatableware, kadomura2013sensing, hori2023masticatory, han2017childish} \\
\rule{0pt}{10pt}
\hspace{1em} Physiological & \\
\rule{0pt}{10pt}
\hspace{2em} Autonomic & \cite{lee2019user, sugita2018diet, kim2016slowee, park2007development, carroll2013food, jung2007wellbeing, nedungadi2018personalized} \\
\rule{0pt}{10pt}
\hspace{2em} Metabolic & \cite{chukwu2011personalized, harous2017hybrid, liang2024exploring, balbin2020scientific, arefeen2024glyman, nedungadi2018personalized} \\
\rule{0pt}{10pt}
\hspace{2em} Cardiac & \cite{huang2018chatbot, park2007development, ludwig2011refresh, jung2007wellbeing, varun2023smart, badawi2012real, nedungadi2018personalized} \\
\rule{0pt}{10pt}
\hspace{1em} Others & \cite{zenun2017online, chaudhry2016evaluation, prasetyo2020foodbot, Burke2011TheEO, Vandelanotte2005EfficacyOS, Haerens2007TheEO, robinson2021humanoid, pels2014fatbelt, rho2017futureself, butscher2016lightweight, mishra2024hybrid, harous2017hybrid, huang2018chatbot, kanjalkar2024ai, sun2020postcard, moses2023investigating, arefeen2022computational, ludwig2011refresh, jung2007wellbeing, bastida2023promoting, nedungadi2018personalized} \\
\rule{0pt}{10pt}
\textbf{Environment-based} & \\
\rule{0pt}{10pt}
\hspace{1em} Visual & \cite{kadomura2013sensing, ganesh2014foodworks, mishra2024hybrid, kanjalkar2024ai, fuchs2019impact, cai2024see, arefeen2022computational, faltaous2021wisdom, nakamura2023eat2pic, kadomura2014persuasive, achilleos2017health, chiu2009playful, vazquez2015development, wang2021research, yr2024nutrispy, alloghani2016mobile, hsiao2011intelligent, fernandez2013virtual, gutierrez2018phara, eigen2018meal, narumi2011augmented, cordeiro2015rethinking, volkova2016smart, waltner2015mango, epstein2016crumbs, eyles2017salt, mankoff2002using, akter2021foodbyte, kwon2016smart} \\
\rule{0pt}{10pt}
\hspace{1em} Acoustic & \cite{faltaous2021wisdom, tommy2017interactive, liu2024hicclip, poddar2024aqua, wijanarko2019fuzzy, ravindran2022hydrationcheck, varun2023smart} \\
\rule{0pt}{10pt}
\hspace{1em} Force & \cite{ritschel2018drink, mitchell2015really, joi2016interactive, hermans2017effect, jung2017mom, zhao2021funeat, krishna2025nutrifit, lessel2016watercoaster, lo2007playful, kim2016ecomeal, ioakimidis2009method, de2020multimodal, maramis2014preventing} \\
\rule{0pt}{10pt}
\hspace{1em} Spectral & \cite{ritschel2018drink, singhal2024multifunctional, bedri2022fitnibble, tommy2017interactive, wijanarko2019fuzzy, matsui2018light, bastida2023promoting, varun2023smart, badawi2012real, kwon2016smart} \\
\rule{0pt}{10pt}
\hspace{1em} Electrical & \\
\rule{0pt}{10pt}
\hspace{2em} Conductivity & \cite{bobin2018smart, tommy2017interactive, kaner2018grow, kreutzer2015base, ko2007mug, pankajavalli2017hydration, bastida2023promoting, varun2023smart, badawi2012real, kwon2016smart} \\
\rule{0pt}{10pt}
\hspace{2em} Magnetic & \cite{vazquez2015development, han2017childish, huang2010context, jung2017mom, yildiz2019wwall} \\
\hline

\end{tabular}
\label{tab:sensing modality paper}
\end{table}

\newpage 

\section{DISTRIBUTION OF PUBLICATIONS IN \delete{FEEDBACK/}INTERVENTION MODALITY}

\begin{table}[htbp]
\centering
\caption{Taxonomy of \delete{Feedback/}Intervention Modality with Involved Publications}
\begin{tabular}{p{0.6\textwidth}p{0.35\textwidth}}
\textbf{\delete{Feedback/}Intervention Modality} & \textbf{References} \\
\hline
\rule{0pt}{10pt}
\hspace{0em} Visual & \\
\rule{0pt}{10pt}
\hspace{1em} Textual & \cite{liang2024exploring,faltaous2021wisdom,zenun2017online,chaudhry2016evaluation,balbin2020scientific,lee2019user,prasetyo2020foodbot,joi2016interactive,Burke2011TheEO,Vandelanotte2005EfficacyOS,Alshurafa2015RecognitionON,Bi2016AutoDietary:AW,Haerens2007TheEO,jung2017mom,de2024social,kim2018smartwatch,rho2017futureself,butscher2016lightweight,mendi2013food,abisha2017embedded,carroll2013food,kumbhare2023low,arefeen2022computational,yr2024nutrispy,mishra2024hybrid,poddar2024aqua,krishna2025nutrifit,singhal2024multifunctional,alloghani2016mobile,harous2017hybrid,hsiao2010smartdiet,gaikwad2024precision,kreutzer2015base,hsiao2011intelligent,huang2018chatbot,maramis2014preventing,dragoni2017semantic,varun2023smart,ludwig2011refresh,achilleos2017health,alexander2017behavioral,kalantarian2014spectrogram,park2007development,bastida2023promoting,kanjalkar2024ai,chukwu2011personalized,terziouglu2023influencing,fernandez2013virtual,qu2014personal,badawi2012diet,arefeen2024glyman,agarwal2024user,sun2020postcard,parra2023enhancing,casas2018food,gutierrez2018phara,kwon2016smart,bedri2022fitnibble,fuchs2019impact,eigen2018meal,badawi2012real,kamachi2023eating,moses2023investigating,ye2016assisting,turner2017byte,hori2023masticatory,lo2007playful,volkova2016smart,waltner2015mango,epstein2016crumbs,eyles2017salt,robinson2020social,kim2016ecomeal,mankoff2002using,nedungadi2018personalized,tirasirichai2018bloom,pankajavalli2017hydration,wijanarko2019fuzzy,tommy2017interactive,ravindran2022hydrationcheck,ioakimidis2009method,vazquez2015development,akter2021foodbyte,jung2007wellbeing,lee2007lifestyle} \\
\rule{0pt}{10pt}
\hspace{1em} Graphical & \cite{cai2024see, liang2024exploring,faltaous2021wisdom,nakamura2023eat2pic,chaudhry2016evaluation,balbin2020scientific,joi2016interactive,kadomura2014persuasive,tang2015introducing,Scisco2011SlowingBR,kadomura2013sensing,zhao2021funeat,kim2018smartwatch,ganesh2014foodworks,kumbhare2023low,yr2024nutrispy,mishra2024hybrid,alloghani2016mobile,hsiao2011intelligent,maramis2014preventing,kalantarian2014spectrogram,bastida2023promoting,kanjalkar2024ai,agarwal2024user,sun2020postcard,parra2023enhancing,gutierrez2018phara,kwon2016smart,yildiz2019wwall,fuchs2019impact,sugita2018diet,turner2017byte,hori2023masticatory,lo2007playful,cordeiro2015rethinking,chiu2009playful,ko2007mug,tirasirichai2018bloom,wijanarko2019fuzzy,zhang2020eat4thought,lee2007lifestyle}\\
\rule{0pt}{10pt}
\hspace{1em} Symbolic & \cite{de2020multimodal,matsui2018light,fortmann2014waterjewel,tang2015introducing,kaner2018grow,kim2016slowee,han2017childish,kim2016eating,kim2018animated,ganesh2014foodworks,ritschel2018drink,bobin2018smart,lessel2016watercoaster,triantafyllidis2019social,mitchell2015really,hermans2017effect,khot2020swan,chiu2009playful,kim2016ecomeal,ravindran2022hydrationcheck,zhou2021mosswater}\\
\rule{0pt}{10pt}
\hspace{0em} Auditory & \cite{xie2023chibo,kadomura2013educatableware,kleinberger2023auditory,cai2024see,kadomura2014persuasive,koizumi2011chewing,robinson2021humanoid,han2017childish,kadomura2013sensing,mayumi2023design,zhao2021funeat,de2024social,liu2024hicclip,ritschel2018drink,mendi2013food,singhal2024multifunctional,varun2023smart,chukwu2011personalized,terziouglu2023influencing,triantafyllidis2019social,agarwal2024user,parra2023enhancing,luo2021foodscrap,chen2022slnom,farooq2017reduction,sugita2018diet,zhang2020eat4thought,robinson2020social,huang2010context,tommy2017interactive,zhou2021mosswater,ioakimidis2009method}\\
\rule{0pt}{10pt}
\hspace{0em} Haptic & \\
\rule{0pt}{10pt}
\hspace{1em} Vibration & \cite{faltaous2021wisdom, wang2021research, kim2016slowee, kim2016eating, chiu2009playful, kim2018smartwatch, varun2023smart, kamachi2023eating, ye2016assisting, hermans2017effect}\\
\rule{0pt}{10pt}
\hspace{1em} Deformation & \cite{chen2022sspoon, pels2014fatbelt, cai2024see, li2018exploring, zhang2019applying, khot2020swan, singhal2024multifunctional}\\
\rule{0pt}{10pt}
\hspace{0em} Olfactory & \cite{cai2024see, narumi2011augmented, li2018exploring}\\
\rule{0pt}{10pt}
\hspace{0em} Physiological & \\
\rule{0pt}{10pt}
\hspace{1em} Electrodermal & \cite{zheng2025alteration, biyani2019intraoral}\\
\hline
\end{tabular}
\label{tab:feedback/intervention modality paper}
\end{table}

\newpage

\section{\modif{MODALITY CATEGORIES IN SENSING PROCESS}}

\begin{table*}[htbp]
    \centering
    \caption{\modif{Descriptions and category of modality terms in sensing process}}
    \resizebox{0.94\textwidth}{!}{%
    \begin{tabular}{>{\columncolor{gray!20}}l p{1.6cm} p{2cm} p{13cm} c}
     \multicolumn{1}{l}{} & \textbf{Modality}  & \textbf{Submodality}  & \textbf{Description} &   \textbf{Paper count}\\
     \hline
     \rule{0pt}{10pt}& Visual & & Human-based visual captures human-derived visual cues to infer the user’s attention, emotional state, or physical actions during ingestion-related behaviors & 8\\
     \cdashline{2-5}[1pt/5pt]
     \rule{0pt}{10pt}&  &Gaze &Gaze monitors the user’s eye movements, fixation points, and visual attention patterns to infer focus, engagement, or intention during ingestion episodes &2 \\
     \rule{0pt}{10pt}&   &Facial 
     
     expression & Facial expression analyzes dynamic changes in facial muscle movements to detect emotional responses, physiological states, or behavioral reactions associated with eating activities & 4\\
     \rule{0pt}{10pt}&   &Body posture& Body posture captures the orientation, alignment, and movement of the user’s body or head to interpret ingestion-related behaviors via camera, such as eating gestures, drinking actions, or body positioning during meals & 3 \\
     \cdashline{2-5}[1pt/1pt]
     \rule{0pt}{10pt}& Auditory &  & Auditory captures human-generated acoustic signals, including speech and non-speech sounds, to infer behavioral states, contextual cues, or ingestion-related activities & 14\\
     \cdashline{2-5}[1pt/5pt]
     \rule{0pt}{10pt}&  & Speech& Speech detects and analyzes verbal expressions, vocalizations, or conversational cues to infer user intentions, social interactions, or eating-related contexts & 7\\
     \rule{0pt}{10pt}&  &Sound & Sound captures non-verbal acoustic signals generated during ingestion activities, such as chewing sounds, swallowing noises, or utensil interactions, to infer ingestive behaviors or consumption events & 8\\
     \cdashline{2-5}[1pt/1pt]
     \rule{0pt}{10pt}& Haptic& & Haptic captures tactile and kinesthetic information through physical interactions with the body, providing insights into touch-based behaviors during ingestion activities & 14 \\
     \cdashline{2-5}[1pt/1pt]
     \rule{0pt}{10pt}& Motion&  & Motion use inertial measurement unit (IMU), bioelectrical sensors, or spectral sensors to capture the movement and dynamic postures of the human body or its parts, enabling the recognition of behaviors, gestures, or ingestion-related activities & 42\\
     \cdashline{2-5}[1pt/5pt]
     \rule{0pt}{10pt}&  & Gesture& Gesture detects voluntary movements of the hands, arms, or other body parts using motion sensors to infer user actions, commands, or contextual cues during ingestion-related behaviors & 37 \\
     \rule{0pt}{10pt}&  & Chewing and Micro-Head Motion & Chewing and micro-head motion monitors rhythmic jaw movements and subtle head or neck dynamics associated with eating, drinking, or swallowing &  5\\
     \cdashline{2-5}[1pt/1pt]
     \rule{0pt}{10pt}& Physiological&  & Physiological captures internal bodily signals related to brain activity, cardiovascular function, and autonomic system responses, enabling the inference of users’ physiological states during ingestion-related activities & 16\\
     \cdashline{2-5}[1pt/5pt]
     \rule{0pt}{10pt}&  & Cardiac& Cardiac enables continuous monitoring of cardiovascular dynamics, including heart rate (HR) and heart rate variability (HRV), typically through electrocardiogram (ECG) or photoplethysmography (PPG) sensors, to assess physiological responses during eating episodes &  7\\
     \rule{0pt}{10pt}&  & Metabolic& Metabolic captures nutritional and biochemical markers such as blood glucose, cholesterol, and body fat rate, typically using biosensors to assess metabolic responses and long-term ingestion-related health status &  6\\
     \rule{0pt}{10pt} \multirow{-34}*{{\rotatebox{90}{\makecell{Human-based}}}}&  & Autonomic&  Autonomic  captures involuntary signals, including electrodermal activity (EDA) and skin temperature, using biosensors to monitor autonomic activity during ingestion&  7\\
     \hline
     \rule{0pt}{10pt} & Visual & & Environment-based visual captures information about food items, environments, or objects through image-based techniques such as classification, segmentation, or object detection &29\\
      \cdashline{2-5}[1pt/1pt]
     \rule{0pt}{10pt} & Acoustic & &  Acoustic detects environmental or object-related features by analyzing sound wave propagation, reflection, or emission patterns& 7\\
      \cdashline{2-5}[1pt/1pt]
     \rule{0pt}{10pt} & Force & & Force measures mechanical pressure or weight changes to infer interactions with food containers, plates, or utensils during ingestion activities  & 13 \\
      \cdashline{2-5}[1pt/1pt]
     \rule{0pt}{10pt} & Spectral & & Spectral analyzes the absorption, reflection, or emission of electromagnetic waves to non-invasively detect material properties such as food composition or temperature & 10\\
      \cdashline{2-5}[1pt/1pt]
     \rule{0pt}{10pt} & Electrical & &  Electrical captures environmental conditions or object characteristics by measuring electrical properties like resistivity, conductivity, or electromagnetic responses& 15 \\
      \cdashline{2-5}[1pt/5pt]
     \rule{0pt}{10pt} &  & Conductivity & Conductivity measures the electrical conductance of liquids or materials to infer properties such as food consistency, liquid presence, or liquid quantity & 10\\
     \rule{0pt}{10pt} \multirow{-12}*{{\rotatebox{90}{\makecell{Environment-based}}}}&  &  Magnetic & Magnetic  sensing detects the presence, identity, or proximity of ingestion-related objects using techniques such as Radio-Frequency Identification (RFID) or Near-Field Communication (NFC) technologies & 5\\
    \hline
    \end{tabular}}
    \label{tab:sensing_modality_table}
\end{table*}

\section{\modif{MODALITY CATEGORIES IN INTERVENTION PROCESS}}

\begin{table*}[htbp]
    \centering
    \caption{\modif{Descriptions and category of modality terms in intervention process}}
    \resizebox{\textwidth}{!}{%
    \begin{tabular}{p{2cm} p{2cm} p{12cm} c}
    \textbf{Modality}  & \textbf{Submodality}  & \textbf{Description} &   \textbf{Paper count}\\
    \hline
    \rule{0pt}{10pt} Visual & Textual& Delivering feedback through text-based information, such as messages, notifications, or instructions& 85\\
    \rule{0pt}{10pt} & Graphical&Providing feedback using graphical elements like charts, progress bars, icons, or visualizations to convey behavioral states or intervention cues & 41\\
    \rule{0pt}{10pt} & Symbolic& Giving visual feedback through abstract or simplified representations, such as symbols, flashing lights, or color changes, to convey states or prompts with minimal cognitive load or linguistic interpretation & 21\\
    \cdashline{1-4}[1pt/1pt]
    
    \rule{0pt}{10pt} Auditory & & Delivering feedback via sound signals, including speech output, alerts, tones, or auditory cues & 32\\
    \cdashline{1-4}[1pt/1pt]
    
    \rule{0pt}{10pt} Haptic & Vibration & Providing feedback through mechanical oscillations delivered by actuators in devices, enabling users to perceive alerts, prompts, or behavioral cues & 10\\
    \rule{0pt}{10pt} & Deformation & Delivering feedback by altering the physical shape, structure, or volume of an interface, allowing users to perceive behavioral cues through tangible disruptions embedded within their natural interaction actuators & 7\\
    \cdashline{1-4}[1pt/1pt]
    
    \rule{0pt}{10pt} Olfactory & & Providing feedback through controlled release or modulation of odors to stimulate the sense of smell, potentially affecting appetite, cravings, or ingestive behavior & 3\\
    \cdashline{1-4}[1pt/1pt]
    
    \rule{0pt}{10pt} Physiological & Electrodermal & Delivering feedback  by modulating skin conductance through electrical stimulation, aiming to influence autonomic nervous system activity and behavioral states during ingestion processes & 2 \\
    
    \hline
    \end{tabular}}
    \label{tab:feedback_intervention_modality_table}
\end{table*}

\newpage

\section{\modif{DESIGN METRIC BETWEEN SENSING MODALITIES AND TARGET FACTORS}}

\begin{table*}[htbp]
\centering
    \caption{Mapping of Sensing Modalities to Target Factors and Behaviors in Ingestion Health}
    \adjustbox{angle=90,center, max width=0.6\textwidth}{
    \begin{tabular}{|c|c|p{1.9cm} |p{1.5cm} |p{1.5cm} |p{1.5cm} |p{1.6cm} |p{1.6cm} |p{1.6cm} |p{1.6cm} |p{1.5cm} |p{1.5cm} |p{1.5cm} |p{1.5cm}| p{1.9cm}| p{1.6cm}|}
    \cline{4-16}
    \multicolumn{3}{c|}{} & \multicolumn{13}{c|}{\textbf{Sensing Modality}}\\
    \cline{4-16}
    \multicolumn{3}{c|}{} & \multicolumn{7}{c|}{\textbf{Human-based}} & \multicolumn{6}{c|}{\textbf{Environment-based}}\\
    \cline{4-16}
    \multicolumn{3}{c|}{} &  &  &  &  & \multicolumn{3}{c|}{\textbf{Physiological}} &  &  &  &  & \multicolumn{2}{c|}{\textbf{Electrical}}\\
    \cline{8-10}
    \cline{15-16}
     \multicolumn{3}{c|}{} & \multirow{-2}*{\makecell{\textbf{Visual}}} & \multirow{-2}*{\makecell{\textbf{Auditory}}} & \multirow{-2}*{\makecell{\textbf{Haptic}}} & \multirow{-2}*{\makecell{\textbf{Motion}}} 
     & \textbf{Cardiac} &\textbf{Metabolic} & \textbf{Autonomic} & \multirow{-2}*{\makecell{\textbf{Visual}}} & \multirow{-2}*{\makecell{\textbf{Acoustic}}} & \multirow{-2}*{\makecell{\textbf{Force}}} & \multirow{-2}*{\makecell{\textbf{Spectral}}} & \textbf{Conductivity} & \textbf{Magnetic}\\
     \cline{1-16}
       & & \textbf{Food Choices} & facial expression \cite{robinson2020social, triantafyllidis2019social}, gaze (VR) \cite{karkar2018virtual}& speech \cite{prasetyo2020foodbot, arefeen2022computational, luo2021foodscrap,robinson2020social}, intake sound \cite{Bi2016AutoDietary:AW, maramis2014preventing, kalantarian2014spectrogram}& touch \cite{cai2024see, robinson2020social}, oral behavior \cite{abisha2017embedded}, body weight \cite{harous2017hybrid, ludwig2011refresh, bastida2023promoting, nedungadi2018personalized}& wrist trajectory \cite{faltaous2021wisdom,dragoni2017semantic, maramis2014preventing}, Object trajectory \cite{kadomura2013educatableware, nakamura2023eat2pic, joi2016interactive, kadomura2014persuasive, kadomura2013sensing, bedri2022fitnibble}, exercise \cite{kumbhare2023low, mishra2024hybrid, harous2017hybrid,dragoni2017semantic,maramis2014preventing, bastida2023promoting, chukwu2011personalized, badawi2012diet,fuchs2019impact,tirasirichai2018bloom, jung2007wellbeing,vazquez2015development}& blood pressure \cite{huang2018chatbot, jung2007wellbeing,ludwig2011refresh,park2007development,nedungadi2018personalized}, heart rate (ECG) \cite{jung2007wellbeing,park2007development,badawi2012real,nedungadi2018personalized}, blood oxygen \cite{park2007development,nedungadi2018personalized}& CGM \cite{liang2024exploring,arefeen2024glyman,nedungadi2018personalized}, BIA \cite{balbin2020scientific,chukwu2011personalized}, cholesterol \cite{harous2017hybrid} & temperature \cite{park2007development, nedungadi2018personalized,jung2007wellbeing} & food picture \cite{faltaous2021wisdom, nakamura2023eat2pic, cai2024see, kadomura2014persuasive, kadomura2013sensing, ganesh2014foodworks, arefeen2022computational, yr2024nutrispy, mishra2024hybrid,alloghani2016mobile, hsiao2011intelligent, hsiao2011intelligent,achilleos2017health,kanjalkar2024ai,kwon2016smart, eigen2018meal, cordeiro2015rethinking, waltner2015mango, epstein2016crumbs,akter2021foodbyte}, Nutrition Label \cite{hsiao2011intelligent, fernandez2013virtual, gutierrez2018phara, fuchs2019impact, volkova2016smart, eyles2017salt,mankoff2002using, vazquez2015development} & food removal \cite{liu2024hicclip}
       & Meals and plate weight \cite{de2020multimodal, joi2016interactive,jung2017mom, zhao2021funeat, ritschel2018drink, krishna2025nutrifit,maramis2014preventing,kim2016ecomeal}&alcohol \cite{matsui2018light}, temperature \cite{bastida2023promoting,kwon2016smart,badawi2012real}, object state \cite{ritschel2018drink}, food distance \cite{bedri2022fitnibble}& humidity \cite{bastida2023promoting,kwon2016smart,badawi2012real}& \\  
       \cline{3-16}
        & & \textbf{Intake Timing} & & intake sound \cite{Bi2016AutoDietary:AW} & button click \cite{fortmann2014waterjewel} & wrist gesture \cite{ye2016assisting}& & CGM \cite{liang2024exploring} & heart rate \cite{varun2023smart} & &liquid level \cite{varun2023smart} & food weight \cite{lessel2016watercoaster}& temperature \cite{varun2023smart}&humidity \cite{varun2023smart} & \\ 
        \cline{3-16}
        & & \textbf{Environment} & & & & & & & & & & & indoor position \cite{faltaous2021wisdom} & & \\ 
        \cline{3-16}
        & \multirow{-12}*{{\rotatebox{90}{\makecell{\textbf{Context Factors}}}}} & \textbf{Activity Around Ingestion} & & & & postprandial exercise \cite{alexander2017behavioral} & & & & & & & & & \\ 
        \cline{2-16}
        & & \textbf{Intake Quantity} & body gesture \cite{xie2023chibo} & & oral behavior \cite{farooq2017reduction}& bite count \cite{Scisco2011SlowingBR}, object trajectory \cite{zhou2021mosswater,li2018exploring}& & & &food picture \cite{chiu2009playful} & leftovers \cite{faltaous2021wisdom}, liquid level \cite{varun2023smart, tommy2017interactive, ravindran2022hydrationcheck}& liquid level \cite{lessel2016watercoaster} & liquid level \cite{poddar2024aqua}& liquid level \cite{kaner2018grow,kreutzer2015base,ko2007mug,pankajavalli2017hydration,wijanarko2019fuzzy}, alcohol \cite{bobin2018smart} & liquid level \cite{yildiz2019wwall}\\ 
        \cline{3-16}
        & & \textbf{Ingestion Pace and Oral Motor Behavior} & facial expression  \cite{kim2018animated} & intake sound \cite{kleinberger2023auditory,koizumi2011chewing,Bi2016AutoDietary:AW,maramis2014preventing,kamachi2023eating,chen2022slnom} & oral behavior \cite{kim2016slowee, biyani2019intraoral, zheng2025alteration}& wrist trajectory \cite{mendi2013food,lee2019user,Scisco2011SlowingBR,kim2016eating,kim2018smartwatch,maramis2014preventing,turner2017byte}, object trajectory \cite{kadomura2013educatableware, nakamura2023eat2pic,biyani2019intraoral, hermans2017effect,zhang2019applying}, skin distance \cite{hori2023masticatory}& & & wrist action (EMG) \cite{lee2019user,kim2016slowee,sugita2018diet}& & & food weight \cite{de2020multimodal,krishna2025nutrifit,lo2007playful,kim2016ecomeal,ioakimidis2009method, mitchell2015really, hermans2017effect}& & & \\ 
        \cline{3-16}
        & & \textbf{Food Awareness or Emotional State} & &intake sound \cite{kleinberger2023auditory, koizumi2011chewing,zhang2020eat4thought}, speech \cite{mayumi2023design, parra2023enhancing,zhang2020eat4thought} & & wrist trajectory \cite{wang2025reflecting}, object trajectory \cite{khot2020swan} & & & emotion state (EDA) \cite{carroll2013food} & food picture (AR) \cite{narumi2011augmented} & & food weight \cite{mitchell2015really} & & & physical contact \cite{han2017childish}\\ 
        \cline{3-16}
        \multirow{-40}*{{\rotatebox{90}{\makecell{\textbf{Target Factors and Behaviors}}}}} &\multirow{-12}*{{\rotatebox{90}{\makecell{\textbf{Process Behaviors}}}}} & \textbf{Safe Eating Practices} & & & & & & & & hygiene \cite{wang2021research} & & & temperature \cite{singhal2024multifunctional}& & \\ 
        \cline{1-16}
        \end{tabular}}
    \label{tab:sensing-behavior}
\end{table*}

\newpage
\section{\modif{DESIGN METRIC BETWEEN INTERVENTION MODALITIES AND TARGET FACTORS}}
\begin{table*}[htbp]
    \centering
    \caption{Mapping of \delete{feedback/}intervention Modalities to Target Factors and Behaviors in Ingestion Health}
    \resizebox{\textwidth}{!}{%
    \begin{tabular}{|c|c|p{2.1cm} |p{2.1cm} |p{2.1cm} |p{2.1cm} |p{2.1cm} |p{2.1cm} |p{2.1cm} |p{2.1cm} |}
    \cline{4-10}
    \multicolumn{3}{c|}{} & \multicolumn{7}{c|}{\textbf{\modif{Intervention} Modality}}\\
    \cline{4-10}
    \multicolumn{3}{c|}{} & \multicolumn{3}{c|}{\textbf{Visual}} &  &  &  &\\
    \cline{4-6}
    \multicolumn{3}{c|}{} & \textbf{Textual} & \textbf{Graphical} & \textbf{Symbolic} & \multirow{-2}*{\makecell{\textbf{Auditory}}} & \multirow{-2}*{\makecell{\textbf{Haptic}}} & \multirow{-2}*{\makecell{\textbf{Olfactory}}} & \multirow{-2}*{\makecell{\textbf{Physiological}}}  \\
    \cline{1-10}
    & & \textbf{Food Choices} & (total 62 pubs) SMS, app/screen message, AR/MR message \cite{gutierrez2018phara,fuchs2019impact} & (total 26 pubs) app/screen graph, AR/MR graph \cite{ganesh2014foodworks, gutierrez2018phara, fuchs2019impact} & LED \cite{de2020multimodal, matsui2018light, ganesh2014foodworks,kim2016ecomeal, ritschel2018drink, triantafyllidis2019social}& microphone speaker \cite{kadomura2013educatableware,cai2024see, kadomura2014persuasive, robinson2021humanoid, kadomura2013sensing,zhao2021funeat,  de2024social, liu2024hicclip,ritschel2018drink, chukwu2011personalized, terziouglu2023influencing, triantafyllidis2019social,agarwal2024user,  luo2021foodscrap, robinson2020social} & piezoelectric vibration \cite{faltaous2021wisdom}, Aerodynamic deformation \cite{pels2014fatbelt}, friction \cite{cai2024see}& Scented liquid \cite{cai2024see}& \\
    \cline{3-10}
    & & \textbf{Intake Timing} & SMS \cite{Bi2016AutoDietary:AW, varun2023smart}, app/screen message \cite{liang2024exploring, faltaous2021wisdom,ye2016assisting}& app/screen graph \cite{liang2024exploring, faltaous2021wisdom}& pneumatic membrane \cite{xie2023chibo}, LED \cite{fortmann2014waterjewel} &microphone speaker \cite{xie2023chibo}, alarm \cite{varun2023smart} & piezoelectric vibration \cite{faltaous2021wisdom}, motor vibration \cite{varun2023smart, ye2016assisting} & LED \cite{lessel2016watercoaster,ravindran2022hydrationcheck} & \\
     \cline{3-10}
    & & \textbf{Environment} & & & & microphone speaker \cite{faltaous2021wisdom} & & & \\
     \cline{3-10}
    & \multirow{-12}*{{\rotatebox{90}{\makecell{\textbf{Context Factors}}}}}& \textbf{Activity Around Ingestion} & SMS \cite{alexander2017behavioral} & & & & & & \\
     \cline{2-10}
    & & \textbf{Intake Quantity} & SMS \cite{poddar2024aqua,kreutzer2015base,moses2023investigating,pankajavalli2017hydration,wijanarko2019fuzzy}, app/screen message \cite{faltaous2021wisdom,chaudhry2016evaluation,chiu2009playful, Alshurafa2015RecognitionON, maramis2014preventing,varun2023smart,tommy2017interactive}, LED message \cite{ravindran2022hydrationcheck}& app/screen graph \cite{faltaous2021wisdom,chaudhry2016evaluation,Scisco2011SlowingBR, maramis2014preventing,varun2023smart,chiu2009playful, ko2007mug,wijanarko2019fuzzy,tommy2017interactive}, projector \cite{yildiz2019wwall} & pneumatic membrane \cite{xie2023chibo}, thermo -chromic \cite{kaner2018grow}, LED \cite{bobin2018smart, lessel2016watercoaster, ravindran2022hydrationcheck}, living-media \cite{zhou2021mosswater}  & microphone speaker \cite{xie2023chibo, farooq2017reduction}, alarm \cite{varun2023smart, tommy2017interactive},  symbolic auditory \cite{zhou2021mosswater} & piezoelectric vibration \cite{faltaous2021wisdom}, motor vibration \cite{li2018exploring,chiu2009playful,varun2023smart},  pneumatic Deformation \cite{chen2022sspoon}  & essential oil \cite{li2018exploring}& \\
     \cline{3-10}
    & & \textbf{Ingestion Pace and Oral Motor Behavior} & SMS \cite{lee2019user,mendi2013food, krishna2025nutrifit,moses2023investigating}, app/screen message \cite{Bi2016AutoDietary:AW,kim2018smartwatch,maramis2014preventing,kamachi2023eating,turner2017byte,hori2023masticatory, lo2007playful,ioakimidis2009method}  &  app/screen graph \cite{nakamura2023eat2pic,Scisco2011SlowingBR,kim2018smartwatch,maramis2014preventing,turner2017byte,hori2023masticatory, lo2007playful}, AR/MR graph \cite{kim2018animated,sugita2018diet} & LED \cite{de2020multimodal,kim2016slowee,kim2016eating,hermans2017effect,kim2016ecomeal}  & chewing sound feedback \cite{kleinberger2023auditory, chen2022slnom}, microphone speaker \cite{ioakimidis2009method,sugita2018diet, mendi2013food,koizumi2011chewing,kadomura2013educatableware}, physical symbolic feedback \cite{mitchell2015really} & piezoelectric vibration \cite{faltaous2021wisdom, zhang2019applying}, motor vibration \cite{kim2016slowee,kim2016eating,kim2018smartwatch,kamachi2023eating, hermans2017effect}& & Intraoral electrical stimulation \cite{biyani2019intraoral, zheng2025alteration}\\
     \cline{3-10}
    & & \textbf{Food Awareness or Emotional State} &SMS\cite {carroll2013food}, app/screen message \cite{mayumi2023design,sun2020postcard,parra2023enhancing,wang2025reflecting}&   app/screen graph\cite{sun2020postcard,parra2023enhancing,zhang2020eat4thought}, AR/MR graph \cite{narumi2011augmented} & LED \cite{han2017childish,khot2020swan} & microphone speaker \cite{kadomura2013educatableware,koizumi2011chewing, han2017childish,mayumi2023design,parra2023enhancing}, chewing sound feedback \cite{kleinberger2023auditory, zhang2020eat4thought}, physical symbolic feedback \cite{mitchell2015really} &motor vibration \cite{khot2020swan} & odor filter \cite{narumi2011augmented}& \\
     \cline{3-10}
    \multirow{-40}*{{\rotatebox{90}{\makecell{\textbf{Target Factors and Behaviors}}}}} & \multirow{-25}*{{\rotatebox{90}{\makecell{\textbf{Process Behaviors}}}}} & \textbf{Safe Eating Practices} & & & & &motor vibration \cite{wang2021research, singhal2024multifunctional} & & \\
     \hline
    \end{tabular}}
    \label{tab:intervention-behavior}
\end{table*}

\end{document}